%% file: main.tex
\documentclass[reprint,onecolumn,nofootinbib]{revtex4-2}
\usepackage{xcolor}
\usepackage{array}
\usepackage{comment}
\usepackage{graphicx}
\usepackage{dcolumn}
\usepackage{bm}
\usepackage{hyperref}
\hypersetup{
    colorlinks=true,
    linkcolor=blue,
    urlcolor=blue,
    citecolor=blue
}
\usepackage{float}
\usepackage{amsmath}
\usepackage{mathtools}
\usepackage{enumitem}
\usepackage{cleveref}
\usepackage{booktabs}
\usepackage{multirow}
\usepackage{ragged2e}
\usepackage{makecell}

\usepackage{listings}
\usepackage{rotating}
\usepackage{xcolor}

\definecolor{codegreen}{rgb}{0,0.6,0}
\definecolor{codegray}{rgb}{0.5,0.5,0.5}
\definecolor{codepurple}{rgb}{0.58,0,0.82}
\definecolor{backcolour}{rgb}{0.95,0.95,0.92}
\lstdefinestyle{mystyle}{
  backgroundcolor=\color{backcolour},   commentstyle=\color{codegreen},
  keywordstyle=\color{magenta},
  numberstyle=\tiny\color{codegray},
  stringstyle=\color{codepurple},
  basicstyle=\ttfamily\scriptsize,
  breakatwhitespace=false,         
  breaklines=true,                 
  captionpos=b,                    
  keepspaces=true,                 
  showspaces=false,                
  showstringspaces=false,
  showtabs=false,
  xleftmargin=0.01\textwidth,
  rulecolor=\color[RGB]{200,200,200},
  frame=bt,
  framextopmargin=2pt,
  framexbottommargin=2pt,
  framexleftmargin=10pt,
  tabsize=4
}
\long\def\/*#1*/{}

\usepackage[normalem]{ulem}

\newcommand{\delete}[1]{}

\usepackage[T1]{fontenc}
\usepackage{lmodern}
\usepackage{amsmath,amssymb,mathtools}
\usepackage{microtype}
\usepackage[most]{tcolorbox}
\usepackage{needspace}
\newcommand{\ket}[1]{\lvert #1\rangle}
\newcommand{\bra}[1]{\langle #1\rvert}

\setlist[itemize]{leftmargin=1.3em,itemsep=0.12em,topsep=0.15em}
\newcommand{\OUTER}{\textsc{Outer}}
\newcommand{\INNER}{\textsc{Inner}}
\newcommand{\ANGLE}{\textsc{Angle}}
\newcommand{\TreeCoeff}{\operatorname{TreeMajoranaCoefficients}}

\newcommand{\Tcost}{\operatorname{Cost}_{\mathrm T}}
\newcommand{\SF}{\mathrm{SF}}
\newcommand{\DOne}{\mathrm{D}_{1}}
\newcommand{\QOne}{\mathrm{Q}_{1}}

\newcommand{\idx}[1]{\{0,\ldots,#1-1\}}

\begin{document}

\title{Leveraging verifiable structure in quantum circuits\\ to find more efficient compilations with agentic AI}

\author{Tanuj Khattar}
\affiliation{Google Quantum AI, Santa Barbara, CA 93111, United States}
\author{Nicholas C. Rubin}
\affiliation{Google Quantum AI, Santa Barbara, CA 
93111, United States}
\author{Craig Gidney}
\affiliation{Google Quantum AI, Santa Barbara, CA 
93111, United States}
\author{Michael Broughton}
\affiliation{Google Quantum AI, Santa Barbara, CA 93111, United States}
\author{Guang Hao Low}
\affiliation{Google Quantum AI, Santa Barbara, CA 
93111, United States}
\author{Ryan Babbush}
\affiliation{Google Quantum AI, Santa Barbara, CA 93111, United States}
\date{\today}

\begin{abstract}
We discuss an effective paradigm of utilizing AI agents to better compile quantum circuits when one can constrain bottleneck subroutines to have some form of classically verifiable structure.
For example, when critical subroutines are composed exclusively of reversible classical logic, phasing operations, and measurement based uncomputation, one can efficiently simulate execution for classical inputs.
This property facilitates efficient certification of approximate circuit synthesis via a form of fuzz testing, provided suitable assumptions on the input distribution.
While this pattern was recently exploited to improve quantum cryptanalysis circuits, here we highlight the flexibility of the broader approach by discussing three important applications. (1) The computation of $1/\sqrt{x}$ that bottlenecks real space simulations of the Coulomb operator such as those important for modeling inertial confinement fusion reactors. (2) The decoding of Reed-Solomon codes which bottlenecks the decoded quantum interferometry algorithm when used to approximately solve the optimal polynomial intersection problem.
And (3) the quantum look-up tables and controlled application of Givens rotations that bottleneck the most performant quantum algorithms for simulating chemistry in second quantization.
We discuss reference implementations of these algorithms and how they can be constrained and verified in the interest of seeding a Kaggle competition designed to accelerate AI-based research into the optimization of their compilation.
\end{abstract}

\maketitle
\tableofcontents
\section{Introduction}
\label{sec:intro}

Compiling a fault-tolerant quantum algorithm into a concrete circuit requires navigating a large design space of arithmetic decompositions, ancilla-management schedules, and precision trade-offs.
Until recently, researchers explored this space by hand, often spending months optimizing a single subroutine and proving its correctness step by step.
Automated circuit optimizers~\cite{kissinger2020reducing,xu2022quartz,ruiz2025alphatensor} help locally, but they preserve exact unitaries on small subcircuits rather than changing the algorithm or its approximations.
AI coding agents~\cite{novikov2025alphaevolve} can automate this search if a fast test harness checks each candidate circuit and scores its resource cost in a closed loop.
For general quantum circuits, checking whether a candidate circuit matches a target unitary is QMA-complete~\cite{janzing2005nonidentity} and intractable on a classical computer above roughly 40 qubits.
Fortunately, in many fault-tolerant quantum algorithms, the subroutines that dominate the cost do not use arbitrary quantum interference internally~\cite{BabbushSpectra,Gidney2019b,Lee2020,rubin2024quantum,gidney2025factor2048,pb2g-j9cw,khattar2025optimizeddqi}.
Instead, they use only reversible classical gates, diagonal phase gates, and measurement-based uncomputation.
This structure lets a classical simulator efficiently check their approximate correctness on sampled computational basis inputs.

Ref.~\cite{babbush2026securingellipticcurvecryptocurrencies} used this method in March 2026 to estimate the cost of breaking 256-bit elliptic curve cryptography on the secp256k1 curve with Shor's algorithm.
To follow responsible disclosure, the authors withheld the explicit circuits.
Instead, they released an open-source simulator and test harness with a zero-knowledge proof that their hidden circuit passed the randomized tests.
Soon after Ref.~\cite{babbush2026securingellipticcurvecryptocurrencies} appeared, the \texttt{ecdsa.fail} challenge took their simulator and test harness and built a public competition for secp256k1 point addition circuits.
The harness checked the approximate correctness of each submission in seconds and ranked it by its qubit and Toffoli counts.
This fast verification let the organizers accept submissions without trusting the participants.
Over 100 participants, most without a quantum computing background, used AI coding agents to further reduce the Toffoli counts by roughly half~\cite{long2026recordreplay}.

In this paper, we apply this workflow to three benchmarks beyond cryptanalysis and use them to seed a Kaggle competition.
We make three main contributions:
\begin{enumerate}
    \item We release \texttt{kickmix}, an open-source C++20 and Python package in Qualtran~\cite{harrigan2024qualtran} that builds and simulates circuits with over $10^8$ gates on one CPU core.
    \item We describe a non-interactive verifier that seeds its test inputs with a salted hash of the submitted circuit.
    We also bound the distance between the implemented and ideal output states by the circuit's failure rate on those inputs.
    \item We define three benchmarks and provide a reference implementation for each of them:
    \begin{itemize}
        \item Phasing by $1/\sqrt{x}$ for first-quantized simulations of Coulomb interactions, such as inertial confinement fusion~\cite{rubin2024quantum}.
        \item Reed-Solomon syndrome decoding for decoded quantum interferometry (DQI) on the optimal polynomial intersection problem~\cite{jordan2025dqi,khattar2025optimizeddqi}.
        \item Quantum lookup tables and Majorana rotations for block encodings of electronic-structure Hamiltonians~\cite{pb2g-j9cw}.
    \end{itemize}
\end{enumerate}
Because classical fuzz testing applies only to kickmix circuits, we restrict each benchmark to the kickmix subroutine that dominates the algorithm's cost.
When that subroutine interacts with non-kickmix parts such as arbitrary state preparation, as in Benchmark III, we price those parts with analytical cost formulas.
This pricing ensures that the benchmark score still reflects the full algorithm.

Section~\ref{sec:kickmix} defines the kickmix gate set, describes the \texttt{kickmix} software package, and derives the state-distance bound from fuzz testing.
Section~\ref{sec:kaggle} describes the background, reference circuit, test distribution, and score for each benchmark.
Section~\ref{sec:outlook} discusses how circuit testing might change as AI agents improve.
The appendices derive the DFTHC block-encoding circuits and the polynomial approximations in the $1/\sqrt{x}$ baseline.

\section{Kickmix circuits and fuzz testing via Fiat-Shamir heuristic}
\label{sec:kickmix}

Optimizing quantum circuits with AI agents requires a fast, reliable way to check whether a candidate circuit works.
For general quantum circuits, classical simulation takes exponential time, and deciding whether two circuits implement the same operation is coNP-complete classically and QMA-complete quantumly~\cite{janzing2005nonidentity}.
Ref.~\cite{babbush2026securingellipticcurvecryptocurrencies} introduced a restricted family of quantum circuits, called \emph{kickmix} circuits, built exclusively from reversible classical gates, diagonal phase operations, and measurement-based uncomputation.
Because every instruction in this family maps any computational basis state to a single computational basis state up to a scalar phase, a classical simulator can run a kickmix circuit on computational basis inputs in time linear in gate count.

Although full quantum algorithms rely on quantum interference across superpositions, many of the subroutines that dominate fault-tolerant resource estimates (such as integer arithmetic, quantum lookup tables, and syndrome decoders) can be written as kickmix circuits~\cite{GidneyAdder,rubin2024quantum,khattar2025optimizeddqi, babbush2026securingellipticcurvecryptocurrencies}.
This structure suggests checking candidate compilations by testing them on randomly sampled classical inputs.
However, doing so requires care: one must check in each problem context that approximate correctness on sampled basis states guarantees high fidelity on the quantum states used by the algorithm.
When this holds, combining classical trajectory simulation with the Fiat-Shamir heuristic~\cite{fiat1986prove,babbush2026securingellipticcurvecryptocurrencies} turns random-input fuzz testing into a public, non-interactive verification protocol that requires no secret test set.
To run this verification loop on subroutines exceeding $10^8$ gates, we release \texttt{kickmix} as an open-source C++20 and Python package inside Qualtran~\cite{harrigan2024qualtran}.
On a single CPU core, \texttt{kickmix} builds circuits at $2\times 10^7$ instructions/s and simulates them at $6.0\times 10^{10}$ instructions/s, constructing an $86\text{M}$-gate Decoded Quantum Interferometry circuit in less than $5\text{ seconds}$ and checking $102{,}400$ syndrome-decoder shots in $150\text{ seconds}$.

\subsection{The kickmix gate set}
\label{sec:kickmix_model}

When a kickmix circuit runs on an $n$-qubit computational basis input $\ket{x}$, no instruction creates a superposition over multiple computational basis states.
A classical simulator therefore needs to store only one bit per qubit wire, one bit per classical register wire, and one scalar output phase.
Specifically, every instruction in the kickmix gate set falls into one of four classes:
\begin{enumerate}
    \item Reversible classical gates (\texttt{X}, \texttt{CX}, \texttt{CCX}, and \texttt{SWAP}): Permute computational basis states by updating the stored qubit bits directly.
    \item Diagonal phase gates (\texttt{Z}, \texttt{CZ}, \texttt{CCZ}, \texttt{Z\_POW}, and \texttt{NEG}): Leave the qubit bits unchanged and update the tracked output phase whenever all control qubits hold $\ket{1}$.
    \item Measurement and reset gates (\texttt{HMR} and \texttt{R}): Measure a qubit in the $X$ basis into a classical bit and reset it to $\ket{0}$ (\texttt{HMR}), or reset a qubit directly to $\ket{0}$ (\texttt{R}).
    \item Classical bit operations (\texttt{BIT\_STORE0}, \texttt{BIT\_STORE1}, and \texttt{BIT\_INVERT}): Set classical register bits to $0$ or $1$, or invert them in place.
\end{enumerate}
Every instruction above can be controlled per gate (\texttt{if b}) or across blocks via a condition stack (\texttt{PUSH\_CONDITION} and \texttt{POP\_CONDITION}), running only when all active classical conditions hold $1$.

The \texttt{HMR} instruction (pronounced ``hammer'') applies a Hadamard gate, measures the target qubit $q_t$ in the $Z$ basis, and resets $q_t$ to $\ket{0}$.
When $q_t$ holds a function $f(x)$ of the remaining qubits, this $X$-basis measurement yields a random outcome $b\in\{0,1\}$ and introduces a phase kickback
\begin{equation}
    \ket{x}\ket{f(x)} \xrightarrow{\texttt{HMR}} (-1)^{b\cdot f(x)} \ket{x}\ket{0}.
    \label{eq:hmr_kickback}
\end{equation}
When $b=1$, classically controlled diagonal gates (such as \texttt{Z}, \texttt{CZ}, \texttt{CCZ}, or \texttt{NEG}) can fix the phase error $(-1)^{f(x)}$ introduced by the $X$-basis measurement.
For example, uncomputing a temporary logical-AND wire $\ket{x_1\land x_2}$ requires only one \texttt{HMR} instruction followed by a classically controlled \texttt{CZ} gate when $b=1$, saving a Toffoli gate~\cite{GidneyAdder}.

However, \texttt{HMR} only works when the resulting phase error can be corrected cheaply.
If a circuit applies \texttt{HMR} to a junk qubit that remains entangled with the system without fixing its phase kickback, classical trajectory simulation fails loudly.
The simulator samples a uniformly random bit $b\in\{0,1\}$, flips the tracked output phase if $q_t\cdot b=1$, and resets $q_t$ to $\ket{0}$.
Without a matching phase fixup, that random sign flip causes the check on sampled inputs to fail with probability $1/2$ whenever $q_t = 1$.
The reset instruction \texttt{R} is intended only for deallocating qubits already uncomputed to $\ket{0}$. If $q_t=1$, the simulator applies a random sign flip, preventing circuits from silently discarding uncleaned junk qubits.

\subsection{The kickmix software package in Qualtran}
\label{sec:kickmix_software}

The \texttt{kickmix} library is implemented in C++20 with Python bindings inside Qualtran~\cite{harrigan2024qualtran}.
Inspired by the bit-packed stabilizer engine of Stim~\cite{gidney2021stim}, \texttt{kickmix} is built around two design principles that let humans and AI agents iterate on $10^8$-gate circuits in seconds.

First, performance drives every layer of circuit construction, serialization, and simulation.
Standard Python circuit libraries allocate a separate heap object for every gate and simulate one input at a time, which becomes prohibitively slow at $10^8$ gates.
Instead, \texttt{kickmix} stores a circuit using a struct-of-arrays representation grouped by operand count: one byte array of opcodes (\texttt{OpType}) alongside separate arrays for $1$-, $2$-, and $3$-qubit arguments, classical target and condition bits, and rotation angles.
Each $1$-byte opcode embeds bit flags indicating which operand arrays it consumes, letting the simulator step sequential pointers through only the active arrays without padding every gate to the worst-case size.
Saving and loading a circuit in binary \texttt{.kmb} format transfers each array between memory and disk as a contiguous binary block, avoiding per-gate text formatting and parsing.
During simulation, \texttt{kickmix} bit-packs $64$ to $512$ independent trajectories across CPU SIMD registers so each instruction updates an entire batch of shots in parallel.
The one exception is \texttt{Z\_POW}: because each trajectory maintains its own $128$-bit fixed-precision phase accumulator rather than a single sign bit, a \texttt{Z\_POW} instruction cannot update the whole batch with a single bitwise SIMD operation.
Instead, it scans the active shot mask across all trajectories in the batch and adds the rotation angle to each active shot sequentially.

Second, expressiveness guides the Python builder API so complex end-to-end algorithms remain compact and readable.
To eliminate the repetitive boilerplate that arises when writing multi-stage fault-tolerant algorithms with measurement-based uncomputation, \texttt{kickmix} provides four abstractions:
\begin{enumerate}
    \item Automatic downgrading: Operations accept qubits, classical bits, and compile-time constants interchangeably, automatically replacing classically controlled gates with lower-cost instructions, dropping no-ops on \texttt{False} controls, and supporting block-scoped classical conditions.
    \item NumPy-style broadcasting: Most primitive gates, measurements, and classical bit operations accept both single wires and register arrays (\texttt{km.array}), broadcasting scalars against registers or operating elementwise across registers, so a single builder call replaces an explicit loop over individual qubits.    
    \item Dynamic workspace allocation: The builder provides methods to allocate and free temporary quantum and classical registers within a configurable qubit budget.
    Subroutines can use these methods to request and release scratch space locally, and freed qubits are handed back out to later allocations.
    \item Library of common subroutines: The builder provides a library of commonly used fault-tolerant subroutines, including temporary-\texttt{AND} gates, QROM table lookups, adders, comparators, finite-field arithmetic, and context-managed unary iteration, with paired compute and measurement-based uncompute variants.
\end{enumerate}

\texttt{kickmix} also includes inspection utilities that generate circuit diagrams and interactive SVG flame charts showing where Toffoli gates and qubits are spent across nested subroutines.

\subsection{Fuzz testing via the Fiat-Shamir heuristic}
\label{sec:fiat_shamir}

To verify a candidate kickmix circuit, we simulate it on computational basis inputs using \texttt{km.Simulator} and compare the resulting bitstrings and phases against a classical reference.
When the input domain is small, we can test every basis state exhaustively to certify exact correctness on all inputs.
When the input domain is too large to enumerate, fuzz testing on a random sample of inputs instead certifies approximate correctness by bounding the probability $\varepsilon$ that the circuit fails on a random input.
Because verification reduces to bounding this failure rate $\varepsilon$, the same procedure naturally accommodates approximations in circuit compilations that are correct on almost all inputs, such as dropping low-weight carry bits, truncating intermediate registers, or bounding iteration counts to save Toffolis~\cite{Gidney2019b,babbush2026securingellipticcurvecryptocurrencies}.

A bound on the per-input failure rate $\varepsilon$ also bounds the error of the circuit on coherent superpositions of inputs.
Fix the outcome string $r\in\{0,1\}^m$ of the $m$ \texttt{HMR} and \texttt{R} instructions in a kickmix circuit.
The circuit then maps each computational basis state to one computational basis state times a phase, and we call this linear map $V_r$.
Each \texttt{HMR} or \texttt{R} instruction with outcome $b$ acts as $(\ket{0}\!\bra{0}+(-1)^b\ket{0}\!\bra{1})/\sqrt{2}$, so the Kraus operators of the circuit are $2^{-m/2}V_r$.
On a basis input, every record $r$ therefore has probability $2^{-m}$, which matches the uniform sampling of the simulator.
If we defer every measurement by copying its outcome into a fresh record qubit, the circuit acts on an input $\ket{\psi}=\sum_x\alpha_x\ket{x}$ as the isometry
\begin{equation}
    W\ket{\psi}=2^{-m/2}\sum_{r}V_r\ket{\psi}\otimes\ket{r}.
\end{equation}

The ideal circuit applies the target unitary $U$ and leaves the record in the state $\ket{+}^{\otimes m}$.
Let $\mathcal{B}_r=\{x : V_r\ket{x}\neq U\ket{x}\}$ be the failing inputs for record $r$, and let $\varepsilon_r=\sum_{x\in\mathcal{B}_r}|\alpha_x|^2$.
Then $\varepsilon=\mathbb{E}_r[\varepsilon_r]$ is the probability that a test shot fails when $x\sim|\alpha_x|^2$ and $r$ is uniform.
For example, suppose that an input fails only because an \texttt{R} or uncorrected \texttt{HMR} gate acts on a qubit in $\ket{1}$.
This input fails when $b=1$ and is exact when $b=0$.
It adds half its weight to $\varepsilon$, and it adds no error to the branches with $b=0$.

We assume that each $V_r$ maps distinct inputs in the support of $\ket{\psi}$ to distinct outputs.
This condition holds when each \texttt{HMR} and \texttt{R} gate acts on a qubit that is a function of the other qubits, as in measurement-based uncomputation.
Each $V_r$ then preserves norms on the support of $\ket{\psi}$ and agrees with $U$ outside $\mathcal{B}_r$.
The triangle inequality gives $\|(V_r-U)\sum_{x\in\mathcal{B}_r}\alpha_x\ket{x}\|\leq 2\sqrt{\varepsilon_r}$ for every $r$.
The record states are orthonormal, so the errors of the branches add in quadrature:
\begin{equation}
    \bigl\|W\ket{\psi}-U\ket{\psi}\otimes\ket{+}^{\otimes m}\bigr\|^2
    =\mathbb{E}_r\Bigl[\Bigl\|(V_r-U)\sum_{x\in\mathcal{B}_r}\alpha_x\ket{x}\Bigr\|^2\Bigr]
    \leq 4\varepsilon.
    \label{eq:branch_robustness}
\end{equation}

Two unit vectors at Euclidean distance $d$ have an overlap of at least $1-d^2/2$.
Hence the purified output has fidelity at least $(1-2\varepsilon)^2$ with $U\ket{\psi}\otimes\ket{+}^{\otimes m}$.
Discarding the record cannot decrease this fidelity.
The physical output state is therefore within trace distance $\sqrt{1-(1-2\varepsilon)^2}\leq 2\sqrt{\varepsilon}$ of $U\ket{\psi}$.
This trace distance also bounds the total-variation distance for any later measurement.
This guarantee requires the verifier to sample test inputs from $|\alpha_x|^2$, so each benchmark in Section~\ref{sec:kaggle} specifies this distribution.\footnote{If the verifier accepts phases within $\tau$ turns, each passing input adds an orthogonal error of norm at most $2\pi\tau|\alpha_x|$. These errors add at most $2\pi\tau$ to the bound. For the tolerances in Section~\ref{sec:kaggle}, this term is negligible.}

To prevent a submission from hard-coding outputs on a fixed test set, our verifier generates challenge inputs via the Fiat-Shamir heuristic~\cite{fiat1986prove,babbush2026securingellipticcurvecryptocurrencies}.
Given a serialized circuit file $C$, the verifier hashes the raw bytes of $C$ with SHA-256 and uses that $256$-bit digest to seed a deterministic pseudorandom generator.
Changing even a single byte of $C$ yields a completely new SHA-256 digest and a fresh set of $N$ challenge inputs.
In a competition setting, we also combine the circuit's SHA-256 digest with a private salt to derive the generator seed so that participants cannot perform seed grinding by repeatedly adding no-op instructions to search for a seed that generates easier test inputs.
This construction makes verification non-interactive, deterministic, and reproducible once the salt is published, without maintaining large secret test sets.

On each shot $i\in\{1,\dots,N\}$, the verifier initializes the input registers to a sampled computational basis state, runs the circuit, and checks three conditions:
\begin{enumerate}
    \item Input and output registers hold their expected classical bitstrings.
    \item Every scratch qubit returns cleanly to $\ket{0}$.
    \item The accumulated output phase matches the target phase within the benchmark tolerance.
\end{enumerate}

If a shot violates any of these conditions, the verifier records a failure.
Suppose a circuit fails on $k$ out of $N$ independent shots, giving an empirical failure rate $\hat{\varepsilon} = k/N$.
If the circuit's true failure rate $\varepsilon$ actually exceeded some threshold $\varepsilon^+ > \hat{\varepsilon}$, the probability of observing $k$ or fewer failures across $N$ shots by chance would be at most
\begin{equation}
    \Pr(\text{at most } k \text{ failures in } N \text{ shots} \mid \varepsilon \geq \varepsilon^+)
    \leq \sum_{j=0}^{k} \binom{N}{j} (\varepsilon^+)^j (1 - \varepsilon^+)^{N-j}.
    \label{eq:binomial_tail}
\end{equation}
Setting the right-hand side of Eq.~\eqref{eq:binomial_tail} equal to a target false-pass probability $\delta = 2^{-128}$ defines the one-sided Clopper-Pearson upper confidence bound $\varepsilon^+$ on the true failure rate $\varepsilon$ at confidence level $1 - 2^{-128}$.
In particular, when a circuit passes all $N$ shots ($k=0$), only the $j=0$ term survives, giving $(1 - \varepsilon^+)^N = \delta$ and $\varepsilon^+ \approx \ln(1/\delta)/N \approx 88.7/N$.
For example, if a circuit passes $N = 10{,}240$ shots with zero failures, the chance that it fails on more than $\varepsilon^+ = 1\%$ of inputs is at most $0.99^{10240} < 2^{-128}$, and passing $N = 102{,}400$ shots certifies $\varepsilon^+ < 0.1\%$ at the same $1 - 2^{-128}$ confidence level.
Depending on the benchmark, the verifier either requires $k=0$ across all $N$ shots or allows $k\geq 0$ failures and uses the certified upper bound $\varepsilon^+$ to compute the repetition overhead in the benchmark score.

\section{Kaggle competition}
\label{sec:kaggle}
We are launching a public competition on Kaggle, an online platform for machine learning and algorithmic challenges, featuring three benchmarks from fusion simulation, decoded quantum interferometry, and quantum chemistry.
Each benchmark isolates the \texttt{kickmix} subroutine that dominates the fault-tolerant cost of the full algorithm at a fixed problem size.
In our own runs with AI coding agents, we have found significant improvements over the previously published baselines, and we expect further improvements remain to be discovered.
For each benchmark, we commit to our best internal run via its SHA-256 hash\footnote{SHA-256 digests of our internal runs:\newline Benchmark~I ($1/\sqrt{x}$): \texttt{4b98f38a52a01fc125da7fc2847f9698abe192519dccf894fd758eccc8f37cea}
\newline Benchmark~II (DQI): \texttt{0701f553a9ac4e8af100f66acfccf2fc82cf5a44c2e5aa765b1d4ace01cbaadb}
\newline Benchmark~III (FeMoCo): \texttt{9b8d070f41dd3b6e10a69dbbae9b047847646d1540de30c7b75799f15692f644}} and will release our full solutions at the end of the competition.
With a \$150{,}000 prize pool, we welcome anyone interested in AI-driven optimization, regardless of quantum computing background, to try new ideas and push these benchmarks further.
Below, we motivate each problem, describe the baseline circuit from prior work, and specify the verification contract and score.

\subsection{Benchmark I: Inverse square roots for Coulomb simulations with relevance to fusion reactor design}

\subsubsection{Problem introduction and importance}
Quantum simulation of matter is one of the primary applications for quantum computers.
This simulation capability is even more important where experimental verification is challenging.
One such example we recently explored was the calculation of stopping power relevant to warm-dense-matter simulation~\cite{rubin2024quantum}.
The calculation of a stopping power is critical to the understanding and design of inertial confinement fusion devices.
The stopping power calculation itself is representative of a broader class of dynamical simulations containing core quantum circuit primitives.
Thus our focus here is on one of the highest cost algorithmic building blocks.
The primitive we focus on in this section is the primary bottleneck for the Trotterized time-evolution for a system represented in a first-quantized representation~\cite{rubin2024quantum} and involves the coherent calculation of the inverse distance between two particles.  

Our goal will be to implement the dynamics of $\eta$-electrons represented on a grid interacting under the Hamiltonian
\begin{align}
H = T + U + V
\end{align}
which involves three terms: 1) $T$ the kinetic energy term for the electron and 2) $U$ the electron-nuclei interaction and 3) $V$ the electron-electron interaction.
For an infinite number of grid points the potential terms $U$ and $V$ are diagonal in the position basis and the kinetic energy is diagonal in the momentum basis.
To evolve a system of electrons interacting under this Hamiltonian by Trotter methods the common strategy is to implement $e^{-i \delta T}$ and $e^{-i \delta (U + V)}$, for a small time value $\delta$, in the basis where the operator is approximately diagonal.
In this work we focus on quantum circuit primitives implementing the potential operator evolution
\begin{align}
U =& \sum_{i=1}^{\eta}\sum_{\ell=1}^{L} \frac{Z_{\ell}}{|r_{i} - R_{\ell}|} = \sum_{i=1}^{\eta}\sum_{\ell=1}^{L} \frac{Z_{\ell}}{\sqrt{(x_{i}-R_{x,\ell})^{2} + (y_{i}-R_{y,\ell})^{2} + (z_{i}-R_{z,\ell})^{2}}} \\
V =& \sum_{i\neq j=1}^{\eta}\frac{1}{|r_{i} - r_{j}|} = \sum_{i\neq j=1}^{\eta}\frac{1}{\sqrt{(x_{i} - x_{j})^{2} + (y_{i} - y_{j})^{2} + (z_{i} - z_{j})^{2}}} 
\end{align}
where we have expanded distance in the Euclidean directions to indicate the main computation needed is the inverse square root of the sum of squared differences in each direction.
Our focus will be the electron-electron computation as this needs to be performed ${\eta}\choose{2}$ times for an $\eta$-particle simulation and is the dominant cost for the total Trotter step.

We use $n_{p}$ bits to represent positions in sign-magnitude form on a grid for the electronic wavefunction
\begin{align}
|\psi\rangle = \sum_{x,y,z=0}^{2^{n_{p}} - 1}c(x,y,z)|x\rangle|y\rangle|z\rangle = \sum_{\boldsymbol{r}}c(\boldsymbol{r})|\boldsymbol{r}\rangle.
\end{align}

The quantum algorithm involves pairwise computation of $V(\mathbf{r}_{i},\mathbf{r}_{j})$ into a register and phasing by $e^{-i\delta V(\mathbf{r}_{i},\mathbf{r}_{j})}$ and uncomputing the $V(\mathbf{r}_{i},\mathbf{r}_{j})$ computation for each pair.
As outlined in Ref.~\cite{rubin2024quantum} performing the computation pairwise instead of for all pairs and summing saves ancilla qubits.
The phasing algorithm looks like the following (reproduced from Ref.~\cite{cody2012faster}, Eqs.~15--17):
\begin{align}
\sum_{\boldsymbol{r}_{1},...,\boldsymbol{r}_{\eta}}& c(\boldsymbol{r}_{1},...,\boldsymbol{r}_{\eta}) | \boldsymbol{r}_{1},...,\boldsymbol{r}_{\eta}\rangle |000...\rangle \\
\rightarrow &\sum_{\boldsymbol{r}_{1},...,\boldsymbol{r}_{\eta}} c(\boldsymbol{r}_{1},...,\boldsymbol{r}_{\eta}) | \boldsymbol{r}_{1},...,\boldsymbol{r}_{\eta}\rangle |V(\boldsymbol{r}_{i},\boldsymbol{r}_{j})\rangle  \label{eq:compute_potential}\\
\rightarrow & \sum_{\boldsymbol{r}_{1},...,\boldsymbol{r}_{\eta}}e^{-iV(\boldsymbol{r}_{i},\boldsymbol{r}_{j})\delta }c(\boldsymbol{r}_{1},...,\boldsymbol{r}_{\eta}) | \boldsymbol{r}_{1},...,\boldsymbol{r}_{\eta}\rangle |V(\boldsymbol{r}_{i},\boldsymbol{r}_{j})\rangle  \\
\rightarrow & \sum_{\boldsymbol{r}_{1},...,\boldsymbol{r}_{\eta}}e^{-iV(\boldsymbol{r}_{i},\boldsymbol{r}_{j})\delta }c(\boldsymbol{r}_{1},...,\boldsymbol{r}_{\eta}) | \boldsymbol{r}_{1},...,\boldsymbol{r}_{\eta}\rangle |000...\rangle
\end{align}

The strategy adopted in Refs.~\cite{Haner2018,rubin2024quantum} follows a similar trick that was used in the video game ``Quake 3'' employing a single step of Newton-Raphson after construction of a high enough quality initial guess.
The C code~\cite{quake3_invsqrt} for this is included below
\begin{lstlisting}
float Q_rsqrt( float number )
{
	long i;
	float x2, y;
	const float threehalfs = 1.5F;

	x2 = number * 0.5F;
	y  = number;
	i  = * ( long * ) &y;						// evil floating point bit level hacking
	i  = 0x5f3759df - ( i >> 1 );               // what the f***?
	y  = * ( float * ) &i;
	y  = y * ( threehalfs - ( x2 * y * y ) );   // 1st iteration
//	y  = y * ( threehalfs - ( x2 * y * y ) );   // 2nd iteration, this can be removed

#ifndef Q3_VM
#ifdef __linux__
	assert( !isnan(y) ); // bk010122 - FPE?
#endif
#endif
	return y;
}
\end{lstlisting}
which produces an approximation that has a maximum relative error of 0.00175228 over the domain of 32 bit floats (IEEE 754-1985).
We work with fixed point representation on quantum computers so we will design an algorithm that works approximately the same but with a fixed point.  

Due to the accuracy requirements of the total time evolution Ref.~\cite{rubin2024quantum} determined that 24-bits of relative error would be needed for the computation in the inverse square root.
Starting from an input register $\ket{x}$ holding the sum of squared coordinate differences, we describe the full algorithm for: 1) normalizing $x$ into a bounded range, 2) computing an initial guess for the inverse square root, and 3) performing one Newton-Raphson iteration followed by phasing and uncomputation.
Attendant to this paper we provide a baseline implementation in Qualtran~\cite{harrigan2024qualtran}.
Our primary task for an AI agent to optimize will be the coherent arithmetic and phasing costs for computing steps 1--3 while maintaining the 24-bit relative error precision bound.
Our baseline implementation is merely a starting point for optimization as there are a number of choices one can make in the computation of the inverse square root.
For example, not using a cubic polynomial over a small rescaled range but a simpler polynomial over a larger range may reduce cost further.
Also the accuracy of coefficients output by QROM can be further optimized.
One could imagine going a completely different route and simply finding a good polynomial approximation over the input domain.
The number of tradeoffs and directions to explore (as in many compilation tasks) makes circuit primitive optimization of this type well posed for AI agent exploration.
Restricting to kickmix circuits allows efficient exploration of these optimizations by many AI agents and will likely lead to large space-time cost reductions.

\subsubsection{Reference implementation}
Calculation of the inverse square-root was sketched in Ref.~\cite{rubin2024quantum} in only enough detail to justify the constant factor analysis.
Here we describe the full implementation which models the classical fixed point method using a single step of Newton-Raphson and provide the reference implementation of the paper.
Computing $w = (x_{i} - x_{j})^{2} + (y_{i} - y_{j})^{2} + (z_{i} - z_{j})^{2}$ from $n_{p}$-bit sign-magnitude coordinates costs $3n_{p}^{2} - n_{p} - 1$ Toffolis (Lemma 8 in Ref.~\cite{Su2021}) and lies upstream of the benchmark subroutine, which takes $\ket{x} = \ket{w}$ directly as its $(2n_{p}+2)$-bit input.
The circuit proceeds in three stages: 1) normalizing $w$ by shifting out pairs of leading zeros, 2) computing an approximate $1/\sqrt{w_{\text{norm}}}$ by polynomial interpolation, and 3) computing a single step of Newton-Raphson iteration followed by phase kickback and uncomputation.
In the fusion resource estimates we used $n_{p}=6$, indicating $w$ has $14$ bits after squaring and lies in the range $\left[1, 11532\right]$; the competition benchmark uses $n_{p}=10$ ($\eta=218$ electrons), giving a $22$-bit input in the range $\left[1, 3(2^{10}-2)^{2}\right] = \left[1, 3{,}133{,}452\right]$.
The range starts at $1$ because we assume an antisymmetrized state as input and thus two electrons cannot occupy the same spatial position on the lattice.
When $w$ is very small $1/\sqrt{w}$ is large-- $w = 1 \mapsto 1/\sqrt{w} = 1$ and $w = 11532 \mapsto 1/\sqrt{w} \approx 0.0093$.
The value $0.0093$ needs $\sim 7$ leading fractional zeros before its first significant bit (and $\sim 11$ leading fractional zeros at $n_{p}=10$).
To get 24 significant bits of relative error at $n_{p}=6$, we would need $24 + 7 = 31$ fractional bits (or $35$ at $n_{p}=10$), making the multiplications expensive.
We address this by shifting and scaling alongside bit truncation during multiplications.

\begin{enumerate}
\item First, Refs.~\cite{Haner2018,rubin2024quantum} avoid the high multiplication cost by normalizing $w$ into a bounded range $[1, 4)$ via bit-shifting.
To see this consider the number of leading zeros in a sampling of $w$ (for $n_{p}=6$):
\begin{align}
\begin{cases}
w = 3     & \mapsto  \texttt{00 00 00 00 00 00 11}     \; \text{(12 leading zeros)} \\
w = 17    & \mapsto  \texttt{00 00 00 00 01 00 01}     \; \text{(8 leading zeros)} \\
w = 4096  & \mapsto  \texttt{01 00 00 00 00 00 00}     \; \text{(1 leading zero)} \\
w = 11532 & \mapsto  \texttt{10 11 01 00 00 11 00}     \; \text{(0 leading zeros)}\\
\end{cases}
\end{align}
all these numbers have different significant figures and thus padding all the values to the appropriate significant figures for $2^{-24}$ is wasteful.
To avoid multiplying many zeros we shift $w$ into a smaller range using the relationship
\begin{align}
w_{\text{norm}} = w \cdot 2^{2s}
\end{align}
where $s$ is the number of leading pairs of zeros.
By using pairs of leading zeros for rescaling we scale our $w \mapsto w_{\text{norm}}$ into the range $w_{\text{norm}} \in \left[1, 4\right)$ so that $1/\sqrt{w_{\text{norm}}} \in \left(0.5, 1\right]$.
Later we will recast the Newton-Raphson step in terms of $w_{\text{norm}}$ and the register storing $s$.
Note that this range implies the first digit after the decimal will always be a $1$.
We need to count and store how many leading zero pairs $s$ there are for each $x$ and produce $x_{\text{norm}}$ by bitshifting.
Each $x$ is stored in a $2n_{p} + 2$ size register ($14$ bits at $n_{p}=6$, $22$ bits at $n_{p}=10$), and we store $s$ in unary in a size $n_{p}$ ancilla register.
To compute the register $|s\rangle$ we sequentially iterate through the bits of $x$ and check if a sequential pair is both zero.
If so, we flip the bit in $s$ corresponding to that pair and shift the register of $x$ forward by $2$ bits.

\item After shifting, $w_{\text{norm}}$ always has at least one of its top 2 bits set, so as a 14-bit integer ($n_{p}=6$): $w_{\text{norm}}  \in \left[4096, 16383\right]$ (i.e., bit 12 or 13 is always 1).
Instead of fitting a polynomial approximating $1/\sqrt{w_{\text{norm}}}$ over $\left[1, 4\right)$ we will fit a polynomial over range $[1, 2]$ and scale appropriately to cover when $x_{\text{norm}} \in \left[2, 4\right)$.

We will split the interval $[1,2]$ in half, and fit the polynomial
\begin{align}\label{eq:a_cubic_poly}
\frac{1}{\sqrt{x}} \approx P_{a}(x) = a_{0} - a_{1}(x - 1) + a_{2}(x - 1)^{2} - a_{3}(x - 1)^{3}
\end{align}
for $x \in [1,3/2]$ and 
\begin{align}\label{eq:b_cubic_poly}
\frac{1}{\sqrt{x}} \approx P_{b}(x) = b_{0} - b_{1}(x - 1.5) + b_{2}(x - 1.5)^{2} - b_{3}(x - 1.5)^{3}
\end{align}
for $x \in [3/2,2]$~\cite{rubin2024quantum} where
\begin{align}
    a_0 &= 0.99994132489119882162, \\
    a_1 &= 0.49609891915903542303, \\
    a_2 &= 0.33261112772430493331, \\
    a_3 &= 0.14876762006038398086.
\end{align}
and
\begin{align}
    b_0 &= 0.81648515205385221995, \\
    b_1 &= 0.27136515484240234115, \\
    b_2 &= 0.12756148214815175348, \\
    b_3 &= 0.044753028579153842218.
\end{align}
In Fig.~\ref{fig:nr_initial_guess} we visualize what this initial guess looks like over the unscaled interval $[1, 2]$ and what happens to the guess after one iteration of Newton-Raphson.
This demonstrates that within double precision one step of Newton-Raphson is sufficient for $2^{-24}$ relative error.
To reuse these polynomials for the $[2,4)$ interval the coefficients must be rescaled.
We describe that derivation in Appendix~\ref{app:polynomial_coeff_rescale}.
\begin{figure}[H]
    \centering
    \includegraphics[width=0.45\linewidth]{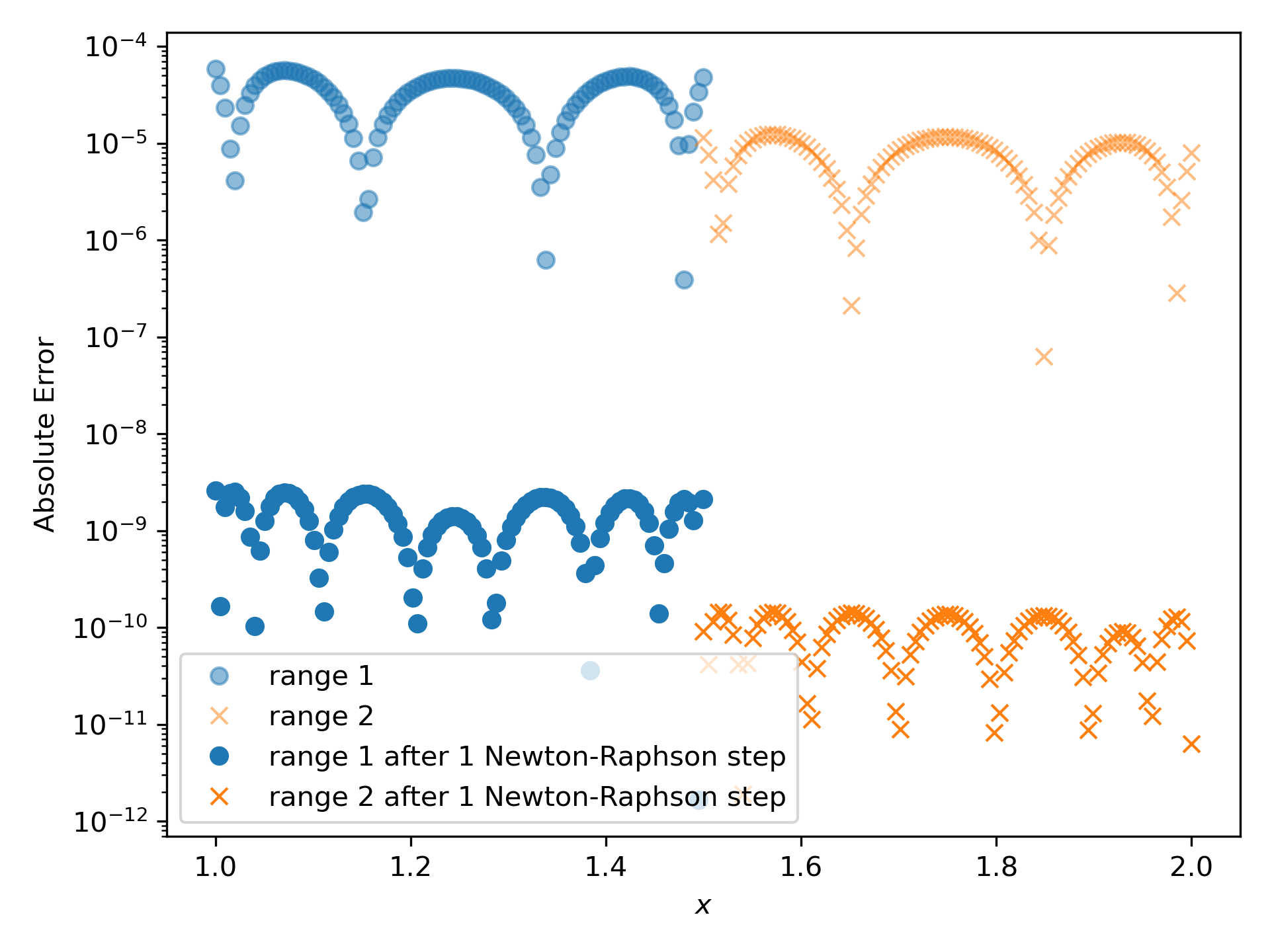}
    \includegraphics[width=0.45\linewidth]{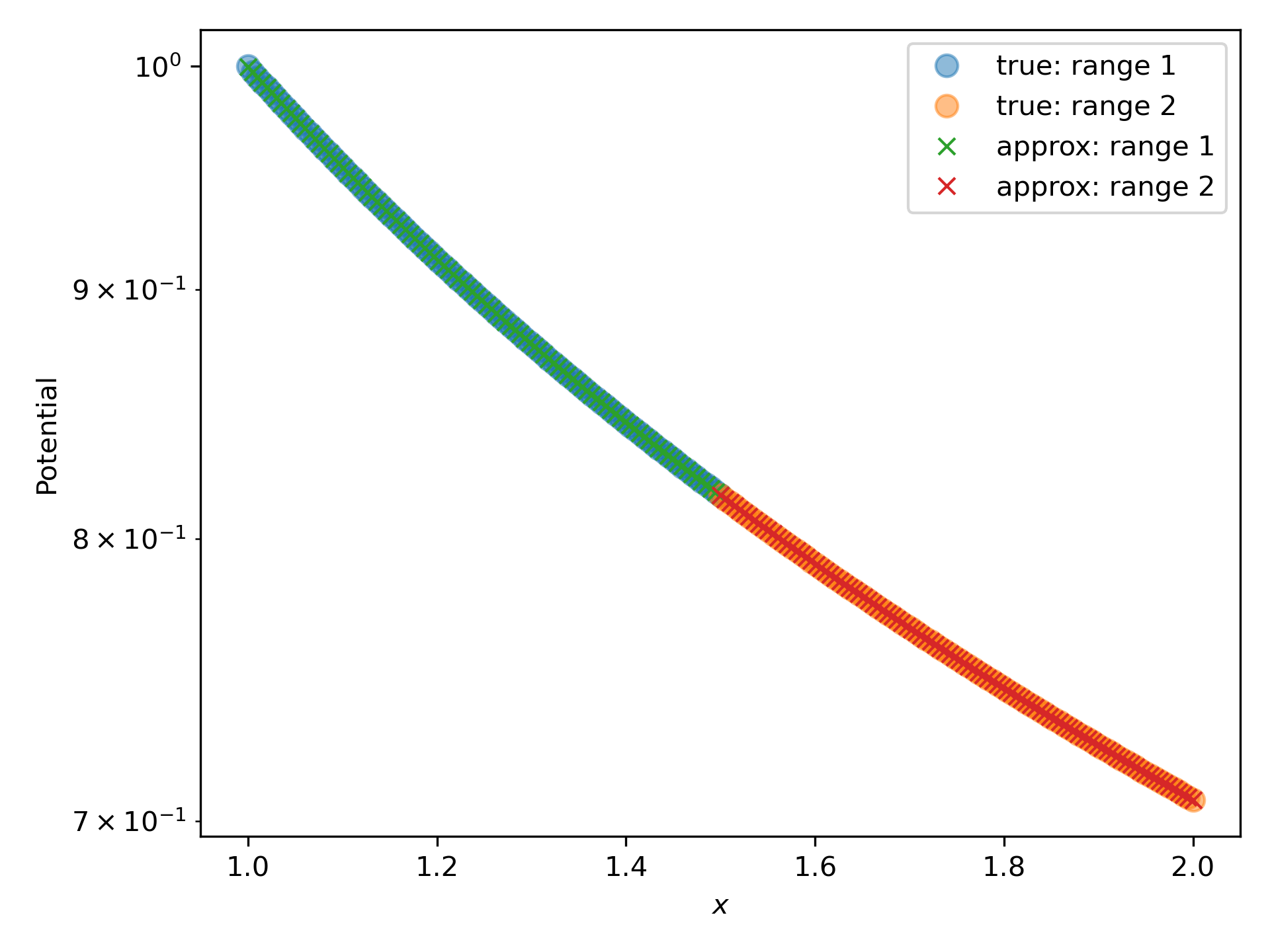}
    \caption{Polynomial approximation on $[1, 2]$ (\texttt{range 1}: $[1, 1.5]$, \texttt{range 2}: $[1.5, 2]$) with $\{a\}$ and $\{b\}$ coefficients from Ref.~\cite{rubin2024quantum}.}
    \label{fig:nr_initial_guess}
\end{figure}

A full description of the algorithmic primitives for QROM output of the multiplication coefficients, how to rescale the coefficients for the $[2,4)$ interval, and the multiplications that need to be performed are provided in Appendix~\ref{app:qrom_poly_approx}.

\item Next, we can see what happens to this $w_{\text{norm}}$ value if we plug it into our Newton-Raphson step.
First we will find an approximate inverse square root to $w_{\text{norm}}$
\begin{align}
y_{0} \approx 1 / \sqrt{w_{\text{norm}}}
\end{align}
which is related to our true inverse square root value
\begin{align}
y_{0} \approx (w \cdot 2^{2s})^{-1/2} = w^{-1/2} \cdot 2^{-s}
\end{align}
if we want to get the ``true'' initial guess we would need to scale $y_{0}$ by $2^{s}$
\begin{align}
y_{\text{init}} = y_{0}\cdot 2^{s} \approx 1/\sqrt{w}
\end{align}
We can now do substitutions into our Newton-Raphson step
\begin{align}
y_{1} =& \frac{1}{2}y_{\text{init}} \left( 3 - w \cdot y_{\text{init}}^{2} \right) \\
=& \frac{1}{2} \left(y_{0} \cdot 2^{s} \right) \left( 3 - w \cdot \left(y_{0} \cdot 2^{s} \right)^{2} \right)   \\
=& \frac{1}{2} \left(y_{0} \cdot 2^{s} \right) \left( 3 - w \cdot 2^{2s} \cdot y_{0}^{2}  \right)   \\ 
=& \frac{1}{2} \left(y_{0} \cdot 2^{s} \right) \left( 3 - w_{\text{norm}} \cdot y_{0}^{2}  \right)   \label{eq:last_nr_mult_xnorm}
\end{align}
which shows we do not need to know the unnormalized $w$ to run Newton-Raphson, we just need to know $w_{\text{norm}}$ and $s$ the number of pair bit shifts.
This rewrite also implies some bitwidth optimizations: $w_{\text{norm}} \cdot y_{0}^{2}$ is very close to $1.0$ and $3 - 1$ will be very close to $2$.
This means for each $w$ we will compute the normalized version by counting leading zero pairs and bit shifting.
The result is $w_{\text{norm}}$ which will go into the infrastructure used to generate $y_{0}$.
During the Newton step we compute $p = \frac{1}{2} y_{0} \left(3 - w_{\text{norm}} \cdot y_{0}^{2}\right)$ with the unshifted 17-bit $y_{0}$ first, and only then apply the unary-controlled left shift $y_{1} = p \ll s$.
Shifting after the multiplication keeps the final multiplier width independent of $n_{p}$, saving $30 n_{p}$ AND gates per direction.
Finally, the circuit phases by one $\texttt{Z\_POW}$ gate per bit of $y_{1}$ and uncomputes stages 1--3 in reverse to restore $\ket{x}\ket{0}$.
\end{enumerate}

$$\ket{x} \xrightarrow{\text{shift by } 2s} x_{\text{norm}} \xrightarrow{\text{QROM + Cubic Poly}} y_{0} \xrightarrow{\text{NR step}} p = \tfrac{1}{2}y_{0}(3 - x_{\text{norm}} y_{0}^{2}) \xrightarrow{\ll s} y_{1} = p \cdot 2^{s} \xrightarrow{\texttt{Z\_POW} \text{ + uncomp.}} e^{2\pi i / \sqrt{x}}\ket{x}$$

For the benchmark size $n_{p} = 10$ ($22$-bit $x$), our reference circuit uses $Q = 260$ qubits, $7{,}735$ Toffoli gates, $1$ $T$ gate, and $37$ non-Clifford $\texttt{Z\_POW}$ rotations, giving a Toffoli-equivalent cost of $C = 7{,}988$, a space-time score of $Q \times C = 2.08 \times 10^{6}$, and a worst-case relative error of $2.84 \times 10^{-8} = 2^{-25.07}$ (at $x = 2{,}588{,}411$).

\subsubsection{Verification strategy}
A submission is a single \texttt{.kmb} circuit file implementing $\ket{x}\ket{0} \mapsto \exp(2\pi i \beta / \sqrt{x})\ket{x}\ket{0}$ for every $x \in [1, \max x]$, where the benchmark default sets $n_{p} = 10$, $\beta = 1$, register width $2n_{p} + 2 = 22$, and $\max x = 3(2^{10}-2)^{2} = 3{,}133{,}452$.
The circuit must declare exactly one named quantum register \texttt{"x"} of $22$ distinct qubits; all remaining qubits are unnamed scratch that must return to $\ket{0}$, and \texttt{"x"} must be restored to $\ket{x}$.
Every input $x \in [1, \max x]$ is simulated $\texttt{shots\_per\_x} = 64$ times ($200{,}540{,}928$ shots in total) to test the randomized measurement outcomes of $\texttt{HMR}$ and $\texttt{R}$ gates during uncomputation, with the simulator PRNG seeded from the SHA-256 digest of the circuit's canonical \texttt{.kmb} binary serialization (combined with a competition salt).
On each shot, the tracked phase $\theta(x)$ (in turns, modulo $1$) is checked in 64-bit fixed point ($\texttt{uint64}$ units of $2^{-64}$ turns) against exact integer square-root arithmetic $\lfloor 2^{65}\beta/\sqrt{x}\rfloor = \texttt{isqrt}(\lfloor 2^{130}\beta^{2}/x\rfloor)$.
Every shot must pass ($k = 0$ failures) within tolerance $\text{tol}(x) = 2^{-24}\beta/\sqrt{x}$ turns.
The benchmark score is $S = Q \times C$, where $Q$ is the circuit's total qubit count and $C = \text{Toffolis} + T/4$ is the mean Toffoli-equivalent cost per shot.
Clifford $\texttt{Z\_POW}$ angles are free, $T$-angles cost $1$ $T$, and each of the $n_{\text{rot}}$ finer $\texttt{Z\_POW}$ rotations is priced analytically via mixed fallback synthesis at $0.56\log_{2}(1/\epsilon) + 5.3$ $T$-gates~\cite{kliuchnikov2023shorter}, where $\epsilon = \pi \min_{x}\bigl(\text{tol}(x) - |\text{error}(x)|\bigr)/n_{\text{rot}}$ splits the remaining phase-error margin evenly across all rotations.

\subsection{Benchmark II: Reed-Solomon decoding for quantum advantage in optimal polynomial intersection}
\label{sec:dqi_benchmark}

\subsubsection{Problem introduction and importance}
Decoded quantum interferometry (DQI) is a quantum algorithm that uses the quantum Fourier transform to reduce optimization problems to decoding problems~\cite{jordan2025dqi}.
We consider its application to the Optimal Polynomial Intersection (OPI) problem.
An OPI instance specifies a set of allowed values $F_i\subseteq\mathbb{F}_q$ at each of the $m=q-1$ nonzero evaluation points $\alpha_i$, for $i=0,\ldots,m-1$.
For $n<m$, we seek a polynomial $Q_x(z)=\sum_{j=0}^{n-1}x_jz^j$ that maximizes
\begin{equation}
    f(x)=\bigl|\{i:Q_x(\alpha_i)\in F_i\}\bigr|.
\end{equation}
Thus, a solution is a low-degree polynomial whose evaluations land in as many of the allowed sets as possible.
The instances considered here have equally sized sets, $|F_i|=r$.

OPI is a promising target for an early demonstration of classically verifiable quantum advantage.
The quality of a sampled solution can be checked efficiently on a classical computer by evaluating $f(x)$, even when finding a comparably good solution is beyond the reach of known classical algorithms analyzed in Refs.~\cite{jordan2025dqi,khattar2025optimizeddqi}.
OPI can be viewed as fitting a low-degree polynomial over a finite field to observations that specify several possible values at each input.
Variants of this problem have been studied as Reed--Solomon list recovery in coding theory~\cite{guruswami2000concatenated} and noisy polynomial reconstruction or interpolation in cryptography~\cite{naor1999oblivious,bleichenbacher2000noisy}.

Let $B_{ij}=\alpha_i^j$ be the $m\times n$ matrix taking polynomial coefficients to evaluations, so that $(Bx)_i=Q_x(\alpha_i)$.
The evaluation vectors $Bx$ form a Reed--Solomon code $C$.
For OPI, DQI uses a syndrome decoder for its dual code $C^\perp=\ker(B^T)$, which has parameters $[m,m-n,n+1]$ and minimum distance $d=n+1$.
If this decoder can correct up to $\ell$ symbol errors, DQI can prepare a state proportional to $\sum_x P(f(x))\ket{x}$ for a polynomial $P$ of degree $\ell$.
Measuring the state samples candidate polynomials $Q_x$ with probability proportional to $|P(f(x))|^2$.
Correcting more errors permits a higher-degree $P$, allowing DQI to sample solutions with a larger expected number of satisfied constraints~\cite{jordan2025dqi}.

The connection between preparing $\sum_x P(f(x))\ket{x}$ and syndrome decoding comes from the quantum Fourier transform (QFT) over $\mathbb{F}_q^n$.
For vectors $s,x\in\mathbb{F}_q^n$ with dot product $s\cdot x=\sum_{j=0}^{n-1}s_jx_j\in\mathbb{F}_q$,
\begin{equation}
    \mathrm{QFT}\ket{s} = \frac{1}{\sqrt{q^n}}\sum_{x\in\mathbb{F}_q^n}\chi(s\cdot x)\ket{x},
    \label{eq:dqi_qft_def}
\end{equation}
where $\chi:\mathbb{F}_q\to\mathbb{C}$ is an additive character satisfying $\chi(u+v)=\chi(u)\chi(v)$.
For $q=2^b$, $\chi(z)=(-1)^{\operatorname{Tr}(z)}\in\{\pm 1\}$, so $\mathrm{QFT}=\mathrm{QFT}^\dagger$ is the Walsh--Hadamard transform.
Write $h_i(z)=\mathbf{1}[z\in F_i]$ and let $b_i\in\mathbb{F}_q^n$ be row $i$ of $B$, so that $Q_x(\alpha_i)=b_i\cdot x$ and $f(x)=\sum_{i=0}^{m-1} h_i(b_i\cdot x)$.
Expanding $h_i:\mathbb{F}_q\to\{0,1\}$ in the 1D Fourier basis of Eq.~\eqref{eq:dqi_qft_def} ($n=1$) gives $h_i(z)=\sum_{y\in\mathbb{F}_q}\hat{h}_i(y)\,\chi(yz)$.
Evaluating at $z=b_i\cdot x$ and pulling the scalar $y\in\mathbb{F}_q$ into the dot product gives
\begin{equation}
    h_i(b_i\cdot x) = \sum_{y\in\mathbb{F}_q} \hat{h}_i(y)\,\chi\bigl(y(b_i\cdot x)\bigr)
    = \sum_{y\in\mathbb{F}_q} \hat{h}_i(y)\,\chi\bigl((y b_i)\cdot x\bigr).
    \label{eq:dqi_single_constraint_fourier}
\end{equation}
Because $h_i(z)\in\{0,1\}$ satisfies $h_i^k=h_i$ for $k\geq1$, expanding $P(f(x))=P\bigl(\sum_{i=0}^{m-1} h_i(b_i\cdot x)\bigr)$ gives a linear combination of products $\prod_{i\in S} h_i(b_i\cdot x)$ over subsets $S\subseteq\{0,\ldots,m-1\}$ of size $|S|\leq\ell$.
Substituting the Fourier expansion~\eqref{eq:dqi_single_constraint_fourier} for each factor $h_i(b_i\cdot x)$ in such a product, with summation variable $y=e_i\in\mathbb{F}_q$ for $i\in S$ and $e_i=0$ for $i\notin S$, defines an error vector $e=(e_0,\ldots,e_{m-1})\in\mathbb{F}_q^m$ of symbol weight $\operatorname{wt}(e)\leq\ell$.
Since $\chi(u)\chi(v)=\chi(u+v)$, multiplying the character factors adds their frequency vectors: $\prod_{i\in S}\chi((e_i b_i)\cdot x) = \chi\bigl((\sum_{i=0}^{m-1} e_i b_i)\cdot x\bigr) = \chi\bigl((B^T e)\cdot x\bigr)$.
Collecting terms and applying Eq.~\eqref{eq:dqi_qft_def} to each mode $\sum_x \chi\bigl((B^Te)\cdot x\bigr)\ket{x}=\sqrt{q^n}\,\mathrm{QFT}\ket{B^Te}$ gives
\begin{equation}
    P(f(x)) = \sum_{\substack{e\in\mathbb{F}_q^m \\ \operatorname{wt}(e)\leq\ell}} a_e\,\chi\bigl((B^T e)\cdot x\bigr),
    \qquad\text{so that}\qquad
    \sum_{x\in\mathbb{F}_q^n} P(f(x))\ket{x}
    = \sqrt{q^n}\,\mathrm{QFT}\!\left(\sum_{\substack{e\in\mathbb{F}_q^m \\ \operatorname{wt}(e)\leq\ell}} a_e\ket{B^T e}\right).
    \label{eq:dqi_fourier_syndrome}
\end{equation}
This is why the polynomial degree $\ell$ and the number of correctable errors coincide: before the final QFT, the basis states $\ket{s}$ are labelled by the syndromes $s=B^Te$ of errors on at most $\ell$ positions.

DQI prepares these components by constructing a superposition of sparse errors and computing their syndromes, giving $\sum_e a_e\ket{e}\ket{B^Te}$.
The amplitudes $a_e$ are determined by $P$ and the Fourier coefficients of the individual constraints.
To obtain the desired Fourier state in the syndrome register, the circuit must erase the error labels without learning which error occurred.
Given $s=B^Te$ and the promise $\operatorname{wt}(e)\leq\ell$, recovering that label is precisely syndrome decoding of $C^\perp$.
When $2\ell\leq n$, the solution is unique: two such errors with the same syndrome would differ by a codeword of weight at most $n$, below the minimum distance $n+1$.
A reversible decoder can therefore erase the error register,
\begin{equation}
    \sum_{\operatorname{wt}(e)\leq\ell}a_e\ket{e}\ket{B^Te}
    \longrightarrow
    \ket{0}\sum_{\operatorname{wt}(e)\leq\ell}a_e\ket{B^Te}.
    \label{eq:dqi_coherent_decode}
\end{equation}
An inverse Fourier transform on the syndrome register then produces the optimization state $\sum_x P(f(x))\ket{x}$, up to normalization.
Our second benchmark targets this decoding computation, which dominates the DQI circuit cost.

The improved construction of Ref.~\cite{khattar2025optimizeddqi} uses measurement-based uncomputation (MBUC) to generate and erase error symbols one at a time.
For $q=2^b$, the circuit reuses a single $b$-qubit error-value register: at position $i$, it prepares $e_i$, adds $e_i b_i$ to the syndrome, then measures the error-value register in the $X$ basis and resets it.
Conditioned on the measurement outcome $c_i\in\{0,1\}^b$, each branch of the superposition labeled by error pattern $e_i$ acquires the sign $(-1)^{c_i\cdot\operatorname{bin}(e_i)}$, determined by the error value stored in that branch.
The accumulated sign from all $m$ error-symbol measurements is therefore $(-1)^{c\cdot\operatorname{bin}(e)}$, where $c=(c_0,\ldots,c_{m-1})$ is the classical measurement record.
Before the final Fourier transform in DQI, the signs introduced by the error-symbol measurements must be cancelled:
\begin{equation}
    \ket{\psi_c}=
    \sum_{\operatorname{wt}(e)\leq\ell}a_e(-1)^{c\cdot\operatorname{bin}(e)}\ket{B^Te}
    \longmapsto
    \sum_{\operatorname{wt}(e)\leq\ell}a_e\ket{B^Te}.
    \label{eq:dqi_measurement_phase}
\end{equation}
Here $c\in\{0,1\}^{mb}$ is the uniformly random measurement record, $\operatorname{bin}(e)$ is the binary encoding of the field symbols in the chosen basis, and the dot product is binary parity modulo two.
Each syndrome $s$ in this superposition uniquely determines a correctable error $e(s)$, so the required sign depends only on $s$ and $c$.

The benchmark is to construct a kickmix circuit that implements this correction on computational-basis inputs:
\begin{equation}
    U_c\ket{s}\ket{0}_W
    =(-1)^{c\cdot\operatorname{bin}(e(s))}\ket{s}\ket{0}_W,
    \qquad s=B^Te(s),\quad \operatorname{wt}(e(s))\leq\ell,
    \label{eq:dqi_benchmark_contract}
\end{equation}
Each test supplies a classical syndrome $s$, encoded as $\ket{s}$, and a classical measurement record $c$; the corresponding error string $e(s)$ is withheld.
The circuit must preserve $s$, return its workspace $W$ to zero, and apply the specified sign, all of which can be checked by classical kickmix simulation.
For fixed $c$, the target operation $U_c$ therefore cancels the signs on the DQI superposition in Eq.~\eqref{eq:dqi_measurement_phase} by linearity.
Implementations may recover error symbols one at a time or compute the required parity directly, allowing AI agents to optimize the decoding algorithm, arithmetic, and cleanup schedule within this contract.

\subsubsection{Reference implementation}
We consider an OPI instance over $\mathbb{F}_{2^b}$ with parameters
\begin{equation}
    (m,n,b,r)=(4095,70,12,2016),
\end{equation}
where $r=|F_i|=(2^b-2^{b/2})/2$ is the bent-function constraint size.
These parameters define a dual Reed--Solomon code with parameters $[4095,4025,71]$ over $\mathbb{F}_{2^{12}}$, for which DQI uses a syndrome decoder correcting up to $\ell=n/2=35$ symbol errors. 
The input syndrome occupies $70\times12=840$ qubits, and the measurement record $c$ contains $4095\times12=49{,}140$ classical bits. 
We represent $\mathbb{F}_{2^{12}}\cong\mathbb{F}_2[\alpha]/(\alpha^{12}+\alpha^6+\alpha^4+\alpha+1)$ in the little-endian polynomial basis $(1,\alpha,\ldots,\alpha^{11})$, order the evaluation points as $\alpha_j=\alpha^j$ for $j=0,\ldots,m-1$, and define syndromes $s_i=\sum_{j=0}^{m-1} e_j \alpha^{(i+1)j}$ for $i=0,\ldots,n-1$. 

The baseline follows the decoder based on the Extended Euclidean Algorithm (EEA) in Ref.~\cite{khattar2025optimizeddqi}.
It has three main stages.
\begin{enumerate}
    \item \textbf{Solve the key equation.} Interpret the syndrome as $S(z)=\sum_{j=0}^{n-1}s_jz^j$.
    Run the EEA on $z^{2\ell}$ and $S(z)$, stopping when the remainder has degree below $\ell$, to obtain an error-locator polynomial $\sigma$ and an error-evaluator polynomial $\Omega$ satisfying the key equation
    \begin{equation}
        \sigma(z)S(z)\equiv\Omega(z)\pmod{z^{2\ell}}.
    \end{equation}
    The error-locator polynomial is proportional to $\prod_{i:e_i\neq0}(1-\alpha_i z)$.
    The zero syndrome is handled by $\sigma=1$ and $\Omega=0$, corresponding to no errors and no phase correction.
    The baseline uses an optimized variant of Proos and Zalka's synchronized EEA construction \cite{proos2004shorsdiscretelogarithmquantum, khattar2025optimizeddqi}, which shares registers between shrinking remainders and growing B\'ezout coefficients and stores temporary quotients in available register space.
    Its output gives explicit access to the B\'ezout coefficients needed by the next stage.

    \item \textbf{Recover symbols and correct their phases.} A Chien search evaluates $\sigma(\alpha_i^{-1})$ at each of the $m$ positions.
    A zero identifies an error location.
    At that location, Forney's formula recovers $e_i$ using $\Omega$ and the formal derivative $\sigma'$.
    The circuit applies $Z$ gates to the recovered symbol bits selected by $c_i$, contributing $(-1)^{c_i\cdot\operatorname{bin}(e_i)}$, then clears the temporary symbol workspace before proceeding.
    Positions without an error contribute no phase.

    \item \textbf{Restore the syndrome and workspace.} Uncompute the EEA from Step~1, restoring the original syndrome and returning all auxiliary registers to zero.
    This preserves the phase applied in Step~2, yielding Eq.~\eqref{eq:dqi_benchmark_contract}.
    
\end{enumerate}

Our reference circuit stores polynomials in two arrays of $n+3$ registers of $b$ qubits each, and reuses the $n$ input syndrome registers in the first array.
Five $O(\log n)$-qubit degree pointers track the active remainders, B\'ezout polynomials, and temporary quotients.
Each of the $3(n+1)+9=222$ synchronized EEA steps uses unary iteration over these pointers to address and shift active polynomial slices.
Over $\mathbb{F}_{2^{12}}$, addition, squaring, and multiplication by fixed powers of $\alpha$ are linear over $\mathbb{F}_2$ and compile to CNOT gates.
Multiplying two quantum field registers uses a $63$-Toffoli Karatsuba circuit, and field inversion uses an Itoh--Tsujii addition chain ($315$ Toffolis and $48$ temporary scaffold qubits).
When both factors of a temporary field product stay in memory, the circuit erases that product by $X$-basis measurement and classically controlled $CZ$ gates~\cite{khattar2025optimizeddqi}.

In Step~2, each of the $m=4095$ positions updates $\sigma$, $\sigma'$, and $\Omega$ in place with CNOT gates and runs one Itoh--Tsujii inversion of $\sigma'(\alpha_i^{-1})$ ($351$ Toffolis per position).
In total, the reference circuit uses $1{,}893$ qubits ($2(n+3)b+O(\log n)$ plus the inversion scaffold) and $5{,}725{,}461$ Toffoli gates after Step~3 cleanup.
This Toffoli count is slightly higher than the $5.2\times 10^6$ decoder count in Ref.~\cite{khattar2025optimizeddqi}, which used a hand-tuned $51$-Toffoli $\mathbb{F}_{2^{12}}$ multiplier instead of our $63$-Toffoli circuit.
Every part of this baseline---explicit remainder storage, pointer-driven EEA steps, polynomial-basis $\mathbb{F}_{2^{12}}$ arithmetic, and per-position Itoh--Tsujii inversion---can be improved.

\subsubsection{Verification strategy}
DQI does not require a decoder that succeeds on every input.
Every candidate polynomial produced by DQI is checked classically by evaluating $f(x)$, so a decoding failure cannot cause an invalid candidate to be accepted and only reduces the probability of sampling a good one.
In the DQI state of Eq.~\eqref{eq:dqi_measurement_phase}, each error $e$ with $\operatorname{wt}(e)\leq\ell$ carries weight $|a_e|^2$ and maps to a distinct syndrome $s=B^Te$.
Let $\varepsilon$ be the probability that the decoder fails on a syndrome $s=B^Te$ with $e\sim|a_e|^2$, averaged over $c$ and its internal measurement outcomes.
By Eq.~\eqref{eq:branch_robustness} and the fidelity argument after it, the output state of the decoder lies within trace distance $2\sqrt{\varepsilon}$ of the ideal state.

Trace distance upper-bounds the total-variation distance between the outcome distributions of any measurement, so the implemented DQI circuit samples candidates within total-variation distance $2\sqrt{\varepsilon}$ of the ideal distribution.
We call a candidate acceptable if it satisfies at least as many constraints as the median output of an ideal DQI run.
An error-free run then succeeds with probability $p_0\geq 1/2$.
A decoder with failure rate $\varepsilon$ succeeds on a single run with probability at least
\begin{equation}
    p\geq \frac{1}{2}-2\sqrt{\varepsilon}.
    \label{eq:dqi_success_bound}
\end{equation}

To test a submission, the verifier uses the seeded generator of Section~\ref{sec:fiat_shamir} to sample an error vector $e\sim|a_e|^2$ with $\operatorname{wt}(e)\leq\ell$.
For the degree-$\ell$ polynomial $P$ maximizing the expected fraction of satisfied constraints (Lemma~9.2 of Ref.~\cite{jordan2025dqi}), the symbol weight $w=\operatorname{wt}(e)\in\{0,\ldots,\ell\}$ has probability $\Pr[\operatorname{wt}(e)=w]=v_w^2$, where $v=(v_0,\ldots,v_\ell)$ is the unit principal eigenvector of the $(\ell+1)\times(\ell+1)$ symmetric tridiagonal matrix $A$ with
\begin{equation}
    A_{w,w}=\frac{w(q-2r)}{\sqrt{r(q-r)}},
    \qquad
    A_{w-1,w}=A_{w,w-1}=\sqrt{w(m-w+1)}.
    \label{eq:dqi_weight_matrix}
\end{equation}
Conditioned on $\operatorname{wt}(e)=w$, the verifier draws $w$ distinct error positions uniformly from $\{0,\ldots,m-1\}$ and draws each nonzero symbol $e_j$ independently and uniformly from $\mathbb{F}_q\setminus\{0\}$ (at these parameters, $\mathbb{E}[\operatorname{wt}(e)]\approx 31.05$ and $\Pr[\operatorname{wt}(e)=35]\approx 2.8\%$).
The verifier then computes the $nb$-bit syndrome $s=B^Te$ and samples the $mb$-bit record $c$ uniformly at random.
The submitted circuit receives $s$ and $c$, but not $e$.
The verifier checks each shot against the three criteria of Section~\ref{sec:fiat_shamir}, with target phase $(-1)^{c\cdot\operatorname{bin}(e)}$.
From $k$ failed shots out of $N=102{,}400$, the verifier computes the upper confidence bound $\varepsilon^{+}$ on $\varepsilon$ at confidence level $1-2^{-128}$ via Eq.~\eqref{eq:binomial_tail}.

The score measures the cost to obtain an acceptable DQI candidate (lower is better).
It weights qubits more heavily than Toffoli gates:
\begin{equation}
    S=R(\varepsilon^{+})\,Q^{2}\,T,
    \qquad
    R(\varepsilon)=\frac{\ln(1-p_0)}{\ln\bigl(1-(p_0-2\sqrt{\varepsilon})\bigr)}
    =\frac{\ln 2}{-\ln\bigl(\tfrac{1}{2}+2\sqrt{\varepsilon}\bigr)},
    \label{eq:dqi_score}
\end{equation}
where we set $p_0=1/2$.
Here $Q$ is the total qubit count (syndrome plus workspace) and $T$ is the single-run Toffoli count, including workspace cleanup.
The factor $Q^2T$ measures the single-run cost, so a 1\% reduction in qubits improves the score as much as a 2\% reduction in Toffoli gates.

The factor $R$ is the repetition overhead from approximate decoding.
If each run succeeds with probability $p$, then $t(p)=\ln\delta'/\ln(1-p)$ independent runs reach an overall failure probability $\delta'$.
Lowering the per-run success probability from $p_0$ to $p_0-2\sqrt{\varepsilon}$ multiplies the number of runs by $t(p_0-2\sqrt{\varepsilon})/t(p_0)=R(\varepsilon)$ for every $\delta'$.
For $\varepsilon^{+}=10^{-5}$, $10^{-4}$, $10^{-3}$, and $10^{-2}$, the repetition factor is $R(\varepsilon^{+})=1.02$, $1.06$, $1.21$, and $1.94$.
Submissions with $\varepsilon^{+}\geq 1/16$ receive no finite score.
With zero failures in $N=102{,}400$ shots, $\varepsilon^{+}\approx 8.7\times10^{-4}$, so every exact submission pays the same floor $R(\varepsilon^{+})\approx1.19$.

\subsection{Benchmark III: QROM and Majorana rotations for quantum simulations of chemistry}

\subsubsection{Problem introduction and importance}
Quantum circuits for state-of-the-art ground-state energy estimation of a second-quantized electronic-structure Hamiltonian involve an important class of data-lookup circuits which are leading Toffoli cost.
These costs are intimately related to the cost of specifying the Hamiltonian and are thus important primitives for algorithms encoding in second quantization.
Thus their compilation is of interest beyond the benchmark provided here. 

Key to many quantum algorithms is the specification of the operator being simulated.
Our primary focus here will be using the linear combination of unitary approach which represents an operator as a sum of unitaries multiplied by a non-negative real weight.
In the chemistry example considered the operator is the Hamiltonian and specifies how the system interacts.
For a system of $n$-qubits the quantum chemistry Hamiltonian can be specified with $\mathcal{O}(n^{4})$ amount of information.
The goal of many modern quantum chemistry algorithms is to figure out how to get the Hamiltonian into the quantum circuit for less than $\mathcal{O}(n^{4})$ costly quantum operations by a combination of classical compression and circuit design. 

Our goal will be to specify the Hamiltonian in Eq.~\eqref{eq:chem_ham} in the lowest space-time volume possible using an LCU where the unitaries are the rank-1 factorizations of the integral tensor.
Almost all modern second quantized simulation techniques leverage compression of the two-body component of Eq.~\eqref{eq:chem_ham} which corresponds to writing the four tensor $h_{pqrs}^{(2)}$ as a tensor network and use a rank-1 unitary representation to build the full operator.
In this benchmark we fix the tensor network to be that of Ref.~\cite{pb2g-j9cw} and require the user to find optimal dimensions of the tensor network that achieve a good approximation of $h_{pqrs}^{(2)}$ and $h_{pq}^{(1)}$ along with a small spacetime volume for the overall circuit.
We cannot ask for the entire compilation of Ref.~\cite{pb2g-j9cw} as that requires a verification framework beyond kickmix, but we will use a proxy to ensure that users cannot artificially change the parameters to lower the space-time cost of the table lookups over the other components of the algorithm.
The combination of classical optimization of the tensor network and how that leads to an efficient quantum circuit is representative of the type of large-scale computational workflow that informs the optimal circuit design.

\subsubsection{Reference implementation}
The block-encoding implementation is factored into self-contained components defined in this section.
The user specifically optimizes over the following components that dominate the overall cost of the algorithm.
\begin{itemize}
\item \cref{sec:DFTHC_factorization}, \cref{eq:dfthc-optimization}: The user explicitly optimizes the DFTHC factorization to Frobenius error $\le\epsilon_{\text{DFTHC}} = 20mHa$, where they have freedom to choose the rank components $(R, B, C)$, the vectors $u_b^{(\rho)}$, and the weights $w_b^{(\rho c)}$ and $W^{(\rho c)}$ (the symmetry shift $(S, \beta_{1})$ is fixed to the provided baseline's), and then minimize cost parameters of block-encoding one-norm and SOS gap $\sqrt{E_\text{gap}(2\Lambda-E_\text{gap})}$. 
\begin{itemize}
\item We provide the variational upper bound on ground state energy.
\item We provide the raw one- and two-electron integrals.
\item We provide a reference factorization that does not necessarily satisfy the accuracy requirements.
\end{itemize}
\item \cref{sec:DFTHC_angles_oracle}, \cref{eq:angle}: The Majorana rotation angles boolean oracle \textsc{Angles}$_j$ is an explicit quantum circuit that the user optimizes over.
The user decides how each batch of the $N-1$ rotation angles is streamed in the $j^\text{th}$ oracle call.
\begin{itemize}
\item We provide a reference quantum circuit compilation of these boolean oracles based on QROM~\cite{BabbushSpectra} quantum lookup table.
\item We provide a reference address to bit-string map used to verify the output of the boolean oracles.
\end{itemize}
\item \cref{sec:Majorana_verification}, \cref{eq:tree_majorana_coefficients}: The user explicitly optimizes the map from rotation angles to Majorana operators by providing a tree specifying the connectivity of Givens rotations. 
\begin{itemize}
\item We provide a reference tree.
\end{itemize}
\end{itemize}
The artifacts are input into the verification procedure of~\cref{sec:DFTHC_block_encoding} to verify a Frobenius error of at most $\epsilon_{\text{DFTHC}}$.

The following are provided to the user as black-boxes with heuristic cost functions.
These are provided as black-boxes as there are too many possible implementations, and most of them would be difficult to verify.
In particular, although many can be decomposed into Boolean oracle calls, these are highly sensitive to the particular implementation and are anyway subdominant in cost.
\begin{itemize}
\item State preparation unitary \textsc{Inner}, heuristic cost function for Toffoli-Qubit spacetime tradeoff (``Inner Prep'' rows of Table 3 in Ref.~\cite{pb2g-j9cw}).
\item Multiplexed state preparation unitary \textsc{Outer}, heuristic cost function for Toffoli-Qubit spacetime tradeoff (``Outer Prep'' rows of Table 3 in Ref.~\cite{pb2g-j9cw}).
\item Majorana Givens rotation circuit, heuristic cost function for Toffolis \cref{sec:Majorana_circuit}.
\end{itemize}

The overall score will be 
\begin{align}
\mathcal{S} = \lambda_{\mathrm{eff}} \times T_{\mathrm{tot}} \times Q_{\mathrm{tot}}, \qquad \lambda_{\mathrm{eff}} = \sqrt{E_\text{gap}(2\Lambda-E_\text{gap})},
\end{align}
where $T_{\mathrm{tot}}$ is the simulated Toffoli count of the $\textsc{Angles}_k$ circuits over one forward and one reverse traversal plus the heuristic Toffoli costs of \textsc{Outer}, \textsc{Inner}, and the $N-1$ Givens rotations, and $Q_{\mathrm{tot}}$ is the number of persistent qubits (the state-preparation registers and the $J b_{\text{rot}}$ angle-word registers) plus the largest temporary workspace among \textsc{Outer}, \textsc{Inner}, and the $\textsc{Angles}_k$ circuits, which run sequentially and share it.
Each $\textsc{Angles}_k$ circuit may use at most $128$ qubits in total; this cap keeps the interesting part of optimizing \textsc{Inner} dominant, and in our previous resource estimates gives a balanced physical cost, e.g.\ 200k physical qubits and $<1$ day.
A submission is rejected if either residual exceeds $\epsilon_{\text{DFTHC}}$, if any $\textsc{Angles}_k$ circuit uses more than $128$ qubits, or if the raw LCU normalization or the SOS gap fail the strict floors $\Lambda > 49.0$\,Ha and $E_{\text{gap}} > 3.914898$\,Ha (see \cref{sec:DFTHC_block_encoding}); these floors rule out degenerate factorizations that shrink the normalization or the gap to game $\lambda_{\mathrm{eff}}$, which itself has no separate floor.
Next we provide detailed descriptions for the numerical optimization and verification steps.

\subsubsection{Optimization of DFTHC factorization}\label{sec:DFTHC_factorization}
\input{GH_notes/dfthc_integrals_optimization_target}

\subsubsection{Verification of circuit for Boolean oracle}\label{sec:DFTHC_angles_oracle}
This is assumed to be a collection of $K$ kickmix circuits that implement the boolean oracle \textsc{Angles}$_k$.
Any ancilla qubits used are assumed to start in, and are returned to, the $\ket{0}$ state.
Each $\textsc{Angles}_k$ oracle must declare named registers `selection' and `word\_0', \ldots, `word\_{J-1}'.
The selection register should be of size $\lceil \log_{2}(N + RB)\rceil$ qubits and take on values $a < N + RB$ in binary.
The first $N$ select the one-body components of the Hamiltonian while the last $RB$ components select the two-particle part.
Each `word\_j' register is $\beta$ qubits wide and holds one rotation angle as a fixed point number in binary; the verifier reads $\beta$ and $J$ from these registers.
The fixed point number is between $[0, 2)$ and thus has a single integer bit and $\beta -1$ mantissa bits.
All other qubits are workspace that must start and end in $\ket{0}$, and \texttt{Z\_POW} instructions are not allowed, so every phase is a multiple of $\pi$.
The family of oracles is simulated for every input in the `selection' register, once forward and once in reverse, with $64$ shots per address, and we verify the construction of the binary representation of angles for Givens rotations.
The user has freedom to specify how the boolean function should work, bits of precision, ordering of the Tree for Givens rotations, and tensor factorizations of the Hamiltonian.
All should be optimized to minimize the cost function heuristic designed to minimize the spacetime footprint of the overall block encoding algorithm.

\subsubsection{Verification of Majorana rotations}\label{sec:Majorana_verification}

\input{GH_notes/majorana_rotations_classical}

\subsubsection{Verification of DFTHC SOSSA block-encoding}\label{sec:DFTHC_block_encoding}
\input{GH_notes/dfthc_block_encoding}

\section{Outlook}
\label{sec:outlook}

The benchmarks in this paper evaluate flat gate streams using logical qubit and Toffoli counts.
While this setup gives AI agents a fast inner loop for subroutine optimization, moving toward end-to-end fault-tolerant compilation suggests two main directions:
\begin{enumerate}
    \item Compiling directly to a target QEC architecture: Logical qubit and Toffoli counts are coarse proxies for physical hardware cost.
    In practice, hardware capabilities dictate the choice of quantum error-correcting (QEC) code.
    That code choice then determines how the architecture implements logical gates and what overheads arise from magic-state distillation, qubit routing, and reaction time~\cite{Gidney2019b,gidney2025factor2048, khattar2025optimizeddqi, zhou2025neutralatom,cain2026shor}.
    Future benchmarks can keep \texttt{kickmix} simulation to check logical correctness while scoring submissions with a QEC cost model for a target architecture.
    AI agents can then co-optimize the logical arithmetic, qubit layout, and gate schedule to minimize physical qubits and runtime.
    \item Machine-checkable proofs of circuit correctness: Classical simulation applies only to kickmix circuits, and a flat \texttt{.kmb} gate stream only tests a single problem size.
    Many quantum subroutines, such as state preparation, quantum signal processing, and fermionic rotations, create superpositions internally and cannot be checked by a kickmix simulator.
    When human researchers optimize these subroutines, they design parameterized constructions for any input size $n$ and prove their correctness and cost bounds symbolically.
    As AI reasoning models and proof assistants such as Lean \cite{deMoura2015} mature, submissions could pair a circuit generator with a proof of its correctness and cost.
    Because a proof assistant checks algebraic identities rather than simulating basis states, it can certify arbitrary quantum unitaries regardless of Hilbert space dimension.
    This shift would let AI agents optimize general quantum algorithms end to end and discover parameterized constructions that hold across all problem sizes.
\end{enumerate}
Together, fast verifiers, architecture-aware cost models, and formal proofs can democratize fault-tolerant compilation, opening quantum algorithm design to researchers and engineers across all fields.

\begin{acknowledgments}
While we were finalizing this manuscript and preparing the competition, Ref.~\cite{gu2026discovering} appeared on the arXiv.
That work also builds on the \texttt{kickmix} model of Ref.~\cite{babbush2026securingellipticcurvecryptocurrencies} to optimize fault-tolerant quantum subroutines with AI agents, and studies many of the same benchmark problems, including Reed-Solomon syndrome decoding for DQI, Coulomb-potential evaluation, and quantum lookup tables for electronic structure.
Although there is significant overlap between the two studies, this work was developed independently and in parallel.
We note that we also spoke about accelerating these benchmarks two weeks prior in a recorded talk at KITP~\cite{babbush2026kitp}.
Our internal results show that all three benchmarks can be further improved beyond the circuits reported to date, so the competition remains open and interesting.
Taken together, the strong results in Ref.~\cite{gu2026discovering} underscore the effectiveness of pairing fast classical verification with AI-driven search, highlighting both the importance of this approach and the relevance of the competition.
\end{acknowledgments}

\raggedbottom
\pagebreak
\bibliography{ref, references}
\appendix
\section{DFTHC block encoding circuits}\label{app:dfthc_circuits}
This appendix expands the tree rotation and the opaque addresses in~\cref{sec:DFTHC_block_encoding} into the labels and select logic used by the DFTHC block encoding.
None of these labels is an optimization variable in the user-facing problem.
\subsection{Tree Majorana coefficients and basis rotation}\label{sec:Majorana_circuit}

Write the ordered edges of $\mathcal T$ as $e_t=(p_t\to q_t)$, $t\in[N-1]$.
For fixed spin $\sigma$, define the edge Givens rotation
\begin{equation}
  G^{(\sigma)}_{pq}(\theta)
  \coloneqq
  \exp\!\left[
    \frac{\theta}{2}
    \left(
      \gamma_{p\sigma0}\gamma_{q\sigma0}
      +\gamma_{p\sigma1}\gamma_{q\sigma1}
    \right)
  \right],
\end{equation}
and let
\begin{equation}
  U_{\mathcal T,\sigma}(\vec\theta)
  \coloneqq
  \prod_{t=0}^{N-2}G^{(\sigma)}_{p_tq_t}(\theta_t),
\end{equation}
with factors written from left to right in increasing $t$.
The classical map $\TreeCoeff$ of~\cref{sec:DFTHC_block_encoding} is computed by
\begin{equation}
\begin{aligned}
  &\mathbf u\leftarrow(1,0,\ldots,0),\\
  &(u_{p_t},u_{q_t})
  \leftarrow
  \bigl(
    \cos\theta_t\,u_{p_t}-\sin\theta_t\,u_{q_t},
    \ \sin\theta_t\,u_{p_t}+\cos\theta_t\,u_{q_t}
  \bigr),
  \qquad t=0,\ldots,N-2.
\end{aligned}
\label{eq:tree-coefficient-recursion}
\end{equation}
It returns a unit vector and satisfies, simultaneously for $x=0,1$,
\begin{equation}
  U_{\mathcal T,\sigma}(\vec\theta)^\dagger
  \gamma_{0\sigma x}
  U_{\mathcal T,\sigma}(\vec\theta)
  =\sum_{p\in[N]}u_p\gamma_{p\sigma x}.
\label{eq:tree-majorana-action}
\end{equation}
Each of the $N-1$ edges is processed once, giving $O(N)$ floating-point operations and $O(N)$ output storage.
Coherently, $\ANGLE_0$ loads the first batch of angle words and each subsequent $\ANGLE_k$ updates that register in place by the XOR difference between adjacent batches.
The corresponding controlled edge rotations are applied in the same tree order.
The fermion-to-qubit representation changes only the fixed Clifford conjugations used to realize each edge rotation; it does not change $\TreeCoeff$, \OUTER, \INNER, or the angle table.

The quantum circuit implementing the $t^\text{th}$ Givens rotation is a $b_t$-bit controlled adder.
The Toffoli count of each adder is assumed to be $2b_t-1$.

\subsection{Sum-of-squares generators}

Let $r\in[R]$, $c\in[C]$, $b\in[B]$, spin
$\sigma\in\{0,1\}$, and generator type
$G\in\{\DOne,\QOne,\SF\}$.
For a real unit vector $u$, define the rotated
Majorana operator
\begin{equation}
  \gamma_{u\sigma x}\coloneqq\sum_{p=0}^{N-1}u_p\gamma_{p\sigma x},
  \qquad x\in\{0,1\}.
\end{equation}
The one-electron positive and negative eigenspaces and the two-electron
squares are represented by
\begin{align}
  O_{\DOne^\sigma,j}
  &\coloneqq
  \sqrt{w_{+,j}}\,
  \frac{\gamma_{u_{+,j}\sigma0}+i\gamma_{u_{+,j}\sigma1}}{2},
  \\
  O_{\QOne^\sigma,j}
  &\coloneqq
  \sqrt{w_{-,j}}\,
  \frac{\gamma_{u_{-,j}\sigma0}-i\gamma_{u_{-,j}\sigma1}}{2},
  \\
  O_{\SF,rc}
  &\coloneqq
  \frac{1}{\sqrt2}
  \left(
    w_B^{(rc)}I
    +\frac{i}{2}\sum_{b\in[B]}\sum_{\sigma\in\{0,1\}}
      w_b^{(rc)}
      \gamma_{u_b^{(r)}\sigma0}\gamma_{u_b^{(r)}\sigma1}
  \right),
\label{eq:generators}
\end{align}
where
$w_B^{(rc)}=W^{(rc)}+\sum_b w_b^{(rc)}$.
Consequently,
\begin{equation}
  H_{\mathrm{DFTHC}}-E_{\mathrm{SOS}}I
  =\sum_{j,\sigma}
      \left(O_{\DOne^\sigma,j}^{\dagger}O_{\DOne^\sigma,j}
      +O_{\QOne^\sigma,j}^{\dagger}O_{\QOne^\sigma,j}\right)
   +\sum_{r,c}O_{\SF,rc}^{\dagger}O_{\SF,rc},
\label{eq:sos-appendix}
\end{equation}
with
\begin{equation}
  E_{\mathrm{SOS}}
  =-2\sum_jw_{-,j}-\frac12\sum_{r,c}|W^{(rc)}|^2.
\end{equation}
The generator normalizations are
\begin{equation}
  \lambda_{\DOne^\sigma,j}=\sqrt{w_{+,j}},\qquad
  \lambda_{\QOne^\sigma,j}=\sqrt{w_{-,j}},\qquad
  \lambda_{\SF,rc}=\frac{1}{\sqrt2}
    \left(|w_B^{(rc)}|+\sum_b|w_b^{(rc)}|\right),
\end{equation}
and $\Lambda$ in Eq.~\eqref{eq:lambda-main} equals one half of the sum of
their squared normalizations, including both spins.

\subsection{Registers, address maps, and select bits}

Let $N_+$ and $N_-$ be the numbers of positive and negative eigenvalues of
$h^{(1)\prime}$, with $N_++N_-=N$.
A contiguous outer address
$x_{\mathrm o}\in[N+RC]$ may be decoded as
\begin{equation}
  x_{\mathrm o}(G,r,c)
  =\begin{cases}
      r, & G=\DOne,\quad r\in[N_+],\\
      N_+ + r, & G=\QOne,\quad r\in[N_-],\\
      N+rC+c, & G=\SF,\quad r\in[R],\ c\in[C].
    \end{cases}
\label{eq:outer-map}
\end{equation}
For a two-electron generator, $b<B$ selects a rotated-Majorana coefficient
and $b=B$ selects its identity coefficient.
A vector address
$\mu\in[N+RB]$ labels the $N$ one-body eigenvectors together with the $RB$
vectors $u_b^{(r)}$.
This is the concrete address denoted abstractly by $a$
in Eq.~\eqref{eq:angle}; the batch label $k$ is fixed in the identity of the oracle $\ANGLE_k$ and is not an input register.

The remaining finite select data comprise: the generator type $G$; separate
spin controls for the outer one-body sum and the inner spin sum in the
$\SF$ generator; the Majorana index $x\in\{0,1\}$ and its relative phase;
the sign of each coefficient; the identity flag $b=B$; and the success and
garbage bits of the chosen state-preparation routine.
These bits contain no
additional continuous Hamiltonian data.

\subsection{Circuit order and cost accounting}

At a high level, one generator is block-encoded by
\begin{equation}
\begin{aligned}
  \OUTER
  &\longrightarrow
  [\text{decode}\to\INNER\to\text{spin select}]
  \longrightarrow
  \{\ANGLE_k\to\text{Givens batch }k\}_{k=0}^{K-1}
  \\
  &\longrightarrow
  \text{Majorana select}
  \longrightarrow
  \{\text{inverse Givens batch }k\to\ANGLE_k\}_{k=K-1}^{0}
  \\
  &\longrightarrow
  [\INNER^{\dagger}\to\text{reflect and square}]
  \longrightarrow
  \OUTER^{\dagger}.
\end{aligned}
\label{eq:pipeline}
\end{equation}
Equation~\eqref{eq:tree-majorana-action} maps the fixed root Majorana to $\gamma_{u\sigma x}$; after this tree rotation the selected Majorana acts on the root spin-orbital in the rotated basis.
A reflection on the inner signal registers and an
inverse application produce the square $O_x^{\dagger}O_x$.
The outer
preparation then forms their weighted sum, yielding Eq.~\eqref{eq:output}.

For any chosen implementation convention, its Toffoli count can be recorded
without exposing the two black boxes as
\begin{equation}
  C_{\mathrm{BE}}
  =\nu_{\mathrm{out}}C_{\mathrm{out}}
   +\nu_{\mathrm{in}}C_{\mathrm{in}}
   +\sum_{k=0}^{K-1}\nu_{\mathrm{ang},k}C_{\mathrm{ang},k}(J)
   +C_{\mathrm{fixed}}(\mathcal T,N,J,b_{\mathrm{rot}}),
\label{eq:cost-accounting}
\end{equation}
where the fixed integers $\nu_{\mathrm{out}},\nu_{\mathrm{in}}$, and $\nu_{\mathrm{ang},k}$ depend only on the selected reflect/uncompute layout.
In the standard forward/select/reverse schedule, every $\ANGLE_k$ is called once in each direction, so the angle-oracle contribution is $2C_{\mathrm{ang}}^{\mathrm{pass}}(J)$.
Only the $C_{\mathrm{ang},k}(J)$ are obtained from explicitly optimized circuits; $C_{\mathrm{out}}$ and $C_{\mathrm{in}}$ remain heuristic inputs.

\section{$1/\sqrt{x}$ baseline circuits}
\subsection{Polynomial rescaling to cover $x_{\text{norm}} \in [2, 4)$}\label{app:polynomial_coeff_rescale}
The range of our interpolation is in $[1,2)$ but the range for $x_{\text{norm}}$ is in $\left[1, 4\right)$.
This suggests that we need some way to relate $x_{\text{norm}} \in \left[2, 4\right)$ back to our interpolation domain of $[1, 2)$.
This is accomplished by noting that we can scale the square root.
Let $u = x_{\text{norm}}/2$
\begin{align}\label{eq:scaling_sqrt}
\frac{1}{\sqrt{x_{\text{norm}}}} = \frac{1}{\sqrt{2u}} = \frac{1}{\sqrt{2}}\frac{1}{\sqrt{u}}
\end{align}
and thus for $x_{\text{norm}} \in \left[2, 4\right)$ we can use scaled coefficients for Eq.~\eqref{eq:a_cubic_poly} and Eq.~\eqref{eq:b_cubic_poly}.
To be more precise about the coefficients we need to use, note that for the upper ranges and by Eq.~\eqref{eq:scaling_sqrt} we need to implement
\begin{align}
\frac{1}{\sqrt{x_{\text{norm}}}} \approx \frac{1}{\sqrt{2}}P(u)
\end{align}
where $P(u)$ is the polynomial in Eq.~\eqref{eq:a_cubic_poly} or Eq.~\eqref{eq:b_cubic_poly}.
To see how we need to modify the coefficients we can plug in our expression for $u=x_{\text{norm}}/2$
\begin{align}\label{eq:rewrite_scaled_p_a}
\frac{1}{\sqrt{x_{\text{norm}}}} =& \frac{1}{\sqrt{2}} \left[ a_0 - a_1\left(\frac{x_{\text{norm}}}{2} - 1\right) + a_2\left(\frac{x_{\text{norm}}}{2} - 1\right)^2 - a_3\left(\frac{x_{\text{norm}}}{2} - 1\right)^3 \right] \\
=& \frac{a_0}{\sqrt{2}} - \frac{a_1}{2\sqrt{2}}(x_{\text{norm}} - 2) + \frac{a_2}{4\sqrt{2}}(x_{\text{norm}} - 2)^2 - \frac{a_3}{8\sqrt{2}}(x_{\text{norm}} - 2)^3
\end{align}
it is noted in Ref.~\cite{rubin2024quantum} that when we scale the square root for some power of two the coefficients need to be scaled as
\begin{enumerate}
\item $a_0 \to a_0 / 2^{m/2}$
\item $a_1 \to a_1 / 2^{3m/2}$
\item $a_2 \to a_2 / 2^{5m/2}$
\item $a_3 \to a_3 / 2^{7m/2}$
\end{enumerate}
with $(x - 1)$ replaced with $(x - 2^m)$.
Plugging in $m=1$ we get
\begin{enumerate}
\item $a_0 / 2^{1/2} \implies a_0 / \sqrt{2}$
\item $a_1 / 2^{3/2} \implies a_1 / 2\sqrt{2}$
\item $a_2 / 2^{5/2} \implies a_2 / 4\sqrt{2}$
\item $a_3 / 2^{7/2} \implies a_3 / 8\sqrt{2}$
\end{enumerate}
 and $(x - 1)$ becomes $(x_{\text{norm}} - 2)$.
 A similar scaling is made for the $b$ coefficients.
 The scaling for the coefficients of the $b$-polynomial follows the same powers of 2 as the $a$-polynomial.
 The only thing that changes is the offset subtracted from $x$.
 Given the interpolation for the upper range ($u \in [1.5, 2]$), the cubic polynomial is
$$P_b(u) = b_0 - b_1(u - 1.5) + b_2(u - 1.5)^2 - b_3(u - 1.5)^3.$$
We want to evaluate $1/\sqrt{x}$, where $x = u \cdot 2^m$ which means $u = \frac{x}{2^m}$.
The true value is $\frac{1}{\sqrt{x}} = \frac{1}{\sqrt{2^m}} \cdot \frac{1}{\sqrt{u}}$ and thus we can approximate it as
$$\frac{1}{\sqrt{x}} \approx \frac{1}{2^{m/2}} \left[ b_0 - b_1\left(\frac{x}{2^m} - 1.5\right) + b_2\left(\frac{x}{2^m} - 1.5\right)^2 - b_3\left(\frac{x}{2^m} - 1.5\right)^3 \right].$$
Now, find a common denominator inside the parentheses.
Notice that $1.5$ is the same as $\frac{1.5 \cdot 2^m}{2^m}$:
$$\left(\frac{x}{2^m} - 1.5\right) = \frac{x - 1.5 \cdot 2^m}{2^m}.$$
Pulling those $2^m$ denominators out of the polynomial terms (squaring and cubing them as we go) gives
$$\frac{1}{\sqrt{x}} \approx \frac{1}{2^{m/2}} \left[ b_0 - \frac{b_1}{2^m}(x - 1.5 \cdot 2^m) + \frac{b_2}{2^{2m}}(x - 1.5 \cdot 2^m)^2 - \frac{b_3}{2^{3m}}(x - 1.5 \cdot 2^m)^3 \right].$$
Finally, we distribute the $\frac{1}{2^{m/2}}$ across all the terms 
$$\frac{1}{\sqrt{x}} \approx \frac{b_0}{2^{m/2}} - \frac{b_1}{2^{3m/2}}(x - 1.5 \cdot 2^m) + \frac{b_2}{2^{5m/2}}(x - 1.5 \cdot 2^m)^2 - \frac{b_3}{2^{7m/2}}(x - 1.5 \cdot 2^m)^3.$$
The scaling factors applied to $b_0, b_1, b_2, b_3$ follow the same rules as the $a$-polynomial:
\begin{enumerate}
 \item   $b_0 \to b_0 / 2^{m/2}$
 \item   $b_1 \to b_1 / 2^{3m/2}$
 \item   $b_2 \to b_2 / 2^{5m/2}$
 \item   $b_3 \to b_3 / 2^{7m/2}$
\end{enumerate}
 And $(x - 1.5)$ is replaced with $(x - 1.5 \cdot 2^m)$.
 \subsection{Example calculation of $|s\rangle$}
 We illustrate the normalization stage on the $n_{p} = 6$ case ($14$-bit register $x_{13}\dots x_{0}$) and give the general cost as a function of $n_{p}$.
 \begin{enumerate}
    \item Starting with the highest two bits ($x_{13}, x_{12}$), we check whether they are \texttt{00} by applying \texttt{X} gates, computing a clean AND into flag $f_{0}$ in the $n_{p}$-qubit unary ancilla register, and uncomputing the \texttt{X} gates.
    \item Controlled on $f_{0}$, we shift the register left by two bits.
    Rather than introducing temporary ancillas for each swap, we split the live bits into their even lane $(x_{0}, x_{2}, \dots, x_{12})$ and odd lane $(x_{1}, x_{3}, \dots, x_{13})$ and apply a controlled left cyclic rotation by 1 position to each lane.
    A controlled cyclic rotation on a lane of $L$ qubits uses $L - 1$ controlled-SWAP (\texttt{CSWAP}) gates, each costing 1 Toffoli in \texttt{kickmix}.
    \item \textbf{Why cyclic rotation leaves clean zeros:} Whenever $f_{0} = 1$, the top pair $(x_{13}, x_{12})$ is guaranteed to be \texttt{00}.
    Rotating the odd lane moves $x_{11} \to x_{13}, \dots, x_{1} \to x_{3}$ and wraps the zero in $x_{13}$ around into $x_{1}$.
    Similarly, rotating the even lane wraps the zero in $x_{12}$ around into $x_{0}$.
    Thus no data collides and the vacated bottom bits are automatically set to \texttt{0} without extra ancilla qubits.
    \item \textbf{Cascading the shifts ($k > 0$):} We repeat this process for $k \in \{1, \dots, n_{p}-1\}$.
    At each step $k$, we check if the new top two bits ($x_{2n_{p}+1}, x_{2n_{p}}$) are \texttt{00}, store the result in flag $f_{k}$, and conditionally rotate the live even and odd bit lanes from bit $2k$ upward left by one position.
    \item \textbf{Truncated Loop Optimization:} Because a shift at step $k$ only occurs if shifts also occurred at all steps $< k$, the bottom $2k$ bits of $x$ were already vacated in previous steps and are guaranteed to be \texttt{0}.
    Each live lane at step $k$ therefore has length $n_{p} + 1 - k$ and needs only $n_{p} - k$ \texttt{CSWAP} gates.
    For $n_{p} = 6$, the number of \texttt{CSWAP} gates across the two lanes decreases linearly:
    \begin{itemize}
        \item Step $k=0$ ($f_0$): two lanes of 7 bits ($x_{13}\dots x_0$) $\rightarrow$ 12 \texttt{CSWAP}s
        \item Step $k=1$ ($f_1$): two lanes of 6 bits ($x_{13}\dots x_2$) $\rightarrow$ 10 \texttt{CSWAP}s
        \item Step $k=2$ ($f_2$): two lanes of 5 bits ($x_{13}\dots x_4$) $\rightarrow$ 8 \texttt{CSWAP}s
        \item Step $k=3$ ($f_3$): two lanes of 4 bits ($x_{13}\dots x_6$) $\rightarrow$ 6 \texttt{CSWAP}s
        \item Step $k=4$ ($f_4$): two lanes of 3 bits ($x_{13}\dots x_8$) $\rightarrow$ 4 \texttt{CSWAP}s
        \item Step $k=5$ ($f_5$): two lanes of 2 bits ($x_{13}\dots x_{10}$) $\rightarrow$ 2 \texttt{CSWAP}s
    \end{itemize}
    Summing over $k = 0, \dots, n_{p}-1$ gives $n_{p}(n_{p}+1)$ \texttt{CSWAP} gates ($42$ at $n_{p}=6$, $110$ at $n_{p}=10$).
    \item \textbf{Total Complexity:} Adding the $n_{p}$ clean AND gates that compute the flags $f_{k}$, forward normalization costs $n_{p}(n_{p}+2)$ Toffoli-equivalent gates ($48$ at $n_{p}=6$, $120$ at $n_{p}=10$).
    During reverse uncomputation at the end of the circuit, the controlled right rotations use $n_{p}(n_{p}+1)$ \texttt{CSWAP} gates while the $n_{p}$ flag ANDs are uncomputed for free by measurement.
    The forward output is the left-aligned register $x_{\text{norm}}$ (whose top two bits are not \texttt{00}) and the $n_{p}$-qubit unary register $|f\rangle$ representing the shift count $s$.
\end{enumerate}

\subsection{Details on cubic polynomial approximation for $1/\sqrt{w}$}\label{app:qrom_poly_approx}
The different polynomial coefficients are input data that must be output coherently via QROM.
The selection state for QROM will be the three most significant bits of $x_{\text{norm}}$.
Recall if $x_{\text{norm}} = b_{13}b_{12}.b_{11}b_{10}...$ with $b_{13}= 1$ then $x_{\text{norm}} \in \left[2, 4\right)$.
We can now work through case-by-case which coefficients to output given we have two polynomials.
\begin{enumerate}
    \item $b_{13}=1$ and $b_{12}=0$ implies that $x_{\text{norm}} \in \left[2, 3\right)$: Output Eq.~\eqref{eq:a_cubic_poly} scaled by something proportional to $1/\sqrt{2}$
    \item $b_{13}=1$ and $b_{12}=1$ implies that $x_{\text{norm}} \in \left[3, 4\right)$: Output Eq.~\eqref{eq:b_cubic_poly} scaled by something proportional to $1/\sqrt{2}$
    \item $b_{13}=0$ (implying $b_{12}=1$) and $b_{11}=0$ implies that $x_{\text{norm}} \in \left[1, 1.5\right)$: Output Eq.~\eqref{eq:a_cubic_poly}
    \item $b_{13}=0$ and $b_{11}=1$ implies that $x_{\text{norm}} \in \left[1.5, 2\right)$: Output Eq.~\eqref{eq:b_cubic_poly}
\end{enumerate}
 The QROM cost for these four entries is $2$ Toffolis because it is not controlled (two AND gates decode $t_{a} = \lnot b_{13} \land b_{11}$ and $t_{b} = b_{13} \land b_{12}$, followed by a CNOT fanout controlled on $\{1, b_{13}, t_{a}, t_{b}\}$ into $|c_{0}\rangle, |c_{1}\rangle, |c_{2}\rangle, |c_{3}\rangle$).

Next, we describe a series of optimizations for evaluating the polynomial once the coefficients are obtained by the QROM read. 
We need to evaluate these polynomials on $x_{\text{norm}}$.
Written as Eq.~\eqref{eq:a_cubic_poly} and Eq.~\eqref{eq:b_cubic_poly} it would seem we need many multiplications and subtractions.
The computation can be significantly simplified by rewriting it with Horner's rule as (for the case of the range $[1,3/2)$)
\begin{align}
    a_0 - (x-1) \{a_1 - (x-1) [a_2 - a_3 (x-1)]\} = a_0 - a_1 (x-1) + a_2 (x-1)^2 - a_3 (x-1)^3.
\end{align}
and for the range $[3/2, 2)$ we have
\begin{align}
 b_0 - (x - 1.5) \{ b_1 - (x - 1.5) [ b_2 - b_3(x - 1.5) ] \} = b_{0} - b_{1}(x - 1.5) + b_{2}(x - 1.5)^{2} - b_{3}(x - 1.5)^{3}.
\end{align}
The above rewrites of the equation need $\Delta x = x-1$ or $\Delta x = x-1.5$ (similarly for our shifted range we need $\Delta x = x-2$ and $\Delta x = x - 3$).
We can derive what shift we need and how it is accomplished by analyzing the polynomials $P_{a}(u)$ and $P_{b}(u)$ across each case:
\begin{enumerate}
\item case 3: $x_{\text{norm}} \in \left[1, 1.5\right)$.
$x_{\text{norm}} = u$ so we use $\Delta x = (x - 1)$.
Binary implementation: $x_{\text{norm}}$ looks like $\texttt{01.0}b_{10} b_{9}\dots$ (since $b_{13}=0, b_{12}=1, b_{11}=0$).
Subtracting $1$ just means flipping $b_{12}$ from \texttt{1} to \texttt{0}.
The $\Delta x$ bits are the fractional bits: $\texttt{00.0}b_{10} b_{9} \dots$.
\item case 4: $x_{\text{norm}} \in \left[1.5, 2\right)$.
Again $x_{\text{norm}} = u$ so we use $\Delta x = (x - 1.5)$.
Binary implementation: $x_{\text{norm}}$ looks like $\texttt{01.1}b_{10} b_{9}\dots$ (since $b_{13}=0, b_{12}=1, b_{11}=1$).
Subtracting $1.5$ means flipping both $b_{12}$ and $b_{11}$ to \texttt{0}.
The $\Delta x$ bits are $\texttt{00.0}b_{10} b_{9} \dots$.
\item case 1: $x_{\text{norm}} \in \left[2, 3\right)$.
For this case $x_{\text{norm}}$ is too large.
We have $u = x_{\text{norm}}/2$ which sets $u \in \left[1, 1.5\right)$.
This means we are using $P_{a}(u)$.
So, 
\begin{align}
(u - 1) = \left( \frac{x_{\text{norm}}}{2} - 1\right) = \frac{1}{2}\left(x_{\text{norm}} - 2\right)
\end{align}
which is the same calculation as Eq.~\eqref{eq:rewrite_scaled_p_a}.
As you can see each of the powers of 2 gets absorbed into a redefinition of the coefficient $a_{i}$.
Binary implementation: $x_{\text{norm}}$ looks like $\texttt{10.}b_{11} b_{10}\dots$ (since $b_{13}=1, b_{12}=0$).
Subtracting $2$ means flipping $b_{13}$ from \texttt{1} to \texttt{0}.
The $\Delta x$ bits are $\texttt{00.}b_{11} b_{10} \dots$.
\item case 2: $x_{\text{norm}} \in \left[3, 4\right)$.
For this case $x_{\text{norm}}$ is too large so we have $u = x_{\text{norm}}/2$ which sets $u \in \left[1.5, 2\right)$.
Following the calculation above we get
\begin{align}
u - 1.5 = \left( \frac{x_{\text{norm}}}{2} - 1.5\right) = \frac{1}{2}\left(x_{\text{norm}} - 3\right)
\end{align}
which matches the scaling coefficients and $\Delta x$ for $P_{b}(u)$ above.
Binary implementation: $x_{\text{norm}}$ looks like $\texttt{11.}b_{11} b_{10}\dots$ (since $b_{13}=1, b_{12}=1$).
Subtracting $3$ means flipping both $b_{13}$ and $b_{12}$ to \texttt{0}.
The $\Delta x$ bits are $\texttt{00.}b_{11} b_{10} \dots$.
\end{enumerate}

We need the bits of $\Delta x$ for the Horner multiplications while leaving $x_{\text{norm}}$ intact for the Newton-Raphson step.
Notice that across all four cases, the integer bits ($b_{13}$ and $b_{12}$) never appear in $\Delta x$, and the lower fractional bits ($b_{10}$ down to $b_0$) from the $0.25$s place downward always equal the corresponding bits of $x_{\text{norm}}$.
Only the $0.5$s-place bit ($b_{11}$) requires controlled logic.
We first examine what $b_{11}$ should be in $\Delta x$ for each of our 4 cases:
\begin{enumerate}
\item[Case 1] ($[2, 3)$): We subtract $2$.
We keep the $0.5$s place.
$\implies$ Keep $b_{11}$.
\item[Case 2] ($[3, 4)$): We subtract $3$.
We keep the $0.5$s place.
$\implies$ Keep $b_{11}$.
\item[Case 3] ($[1, 1.5)$): We subtract $1$.
$b_{11}$ is already \texttt{0}, so $\Delta x_{11} = 0$.
\item[Case 4] ($[1.5, 2)$): We subtract $1.5$, clearing the $0.5$s place.
Even though $b_{11}=1$ in $x_{\text{norm}}$, $\Delta x_{11}$ must be \texttt{0}.
\end{enumerate}
Look at the logic for $b_{11}$ above.
We keep $b_{11}$ in Cases 1 and 2 ($b_{13} = 1$), and clear it in Case 4 ($b_{13} = 0$).
This means $\Delta x_{11} = b_{13} \land b_{11}$.
Rather than allocating a separate $|\Delta x\rangle$ register and copying the lower fractional bits with CNOTs, our circuit computes only this single $0.5$s-place bit $b_{13} \land b_{11}$ into one ancilla qubit using $1$ AND gate, and reads the remaining fractional bits of $\Delta x$ directly from $x_{\text{norm}}$.

Finally, we perform the Horner multiplications and the Newton-Raphson step.
The polynomial to evaluate is
\begin{align}
 y_0 = a_0 - a_1 \Delta x + a_2 \Delta x^2 - a_3 \Delta x^3 = a_0 - \Delta x \cdot \Big(a_1 - \Delta x \cdot (a_2 - a_3 \cdot \Delta x)\Big),
\end{align}
which requires three multiplications and three subtractions.
In our reference circuit, we use $14$ bits of $\Delta x$ (all $12$ fractional bits at $n_{p}=6$, or $14$ of the $20$ fractional bits at $n_{p}=10$), $17$ fractional bits for the QROM coefficients $|c_{0}\rangle, |c_{1}\rangle, |c_{2}\rangle, |c_{3}\rangle$ and Horner intermediates, $28$ fractional bits for $y_{0}^{2}$, $28$ fractional bits plus $2$ integer bits ($30$ qubits) for $z = x_{\text{norm}} y_{0}^{2}$ and $3 - z$, $29$ fractional bits plus $1$ integer bit ($30$ qubits) for $p = y_{0}(3 - z)$, and $30 + n_{p}$ qubits ($40$ at $n_{p}=10$, $36$ at $n_{p}=6$) for $y_{1} = p \ll s$.
A single shared $30$-qubit scratch register holds row partial products across all multipliers.
The remaining arithmetic stages proceed as follows:
\begin{itemize}
    \item \textbf{In-place Horner evaluation:} Rather than allocating six fresh output registers, the circuit updates the coefficient registers in place over three steps: $c_{2} \mathrel{-}= T(c_{3}\Delta x)$, $c_{1} \mathrel{-}= T(c_{2}\Delta x)$, and $c_{0} \mathrel{-}= T(c_{1}\Delta x)$, leaving $y_{0}$ in $|c_{0}\rangle$ ($1{,}071$ Toffolis at $n_{p}=10$, $954$ at $n_{p}=6$).
    Here $T(\cdot)$ drops every partial-product bit below the target register's least significant bit.
    Each bit of $\Delta x$ controls a row update that ANDs the live multiplicand slice into the shared scratch register, subtracts it in place, and immediately uncomputes the row ANDs by measurement.
    \item \textbf{Squaring:} Computes $T(y_{0}^{2})$ into a $28$-bit register ($442$ Toffolis).
    Diagonal bits $y_{k}^{2} = y_{k}$ ($2k \le 28$) are copied with CNOTs, and doubled cross terms $2y_{k}y_{l}$ ($k + l - 1 \le 28$) are added row by row.
    \item \textbf{Mantissa multiply and correction:} Multiplies $T(y_{0}^{2})$ by all $2n_{p}+2$ bits of $x_{\text{norm}}$ into the $30$-qubit register $z = T(x_{\text{norm}} \cdot T(y_{0}^{2}))$ ($1{,}043$ Toffolis at $n_{p}=10$, $719$ at $n_{p}=6$), then replaces $z$ in place with $3 - z$ in two's complement by applying bitwise \texttt{X} gates and adding $(3 \ll 28) + 1$ ($28$ Toffolis).
    \item \textbf{Final NR multiply and denormalization:} Computes $p = T(y_{0} \cdot z) \approx 2/\sqrt{x_{\text{norm}}}$ into the bottom $30$ qubits of $y_{1}$ ($867$ Toffolis), then shifts $y_{1} = p \ll s$ in place by applying a controlled cyclic left shift of $1$ position on the $31$-qubit slice $y_{1}[k:k+31]$ controlled on $f_{k}$ for $k = 0, \dots, n_{p}-1$ ($30 n_{p}$ \texttt{CSWAP} gates: $300$ at $n_{p}=10$, $180$ at $n_{p}=6$).
    \item \textbf{Phasing and uncomputation:} Applies one $\texttt{Z\_POW}$ gate of $2^{j - F}$ turns on each bit $j \in \{0, \dots, F-1\}$ of $y_{1}$ ($F = 30 + n_{p}$), and then runs all stages in reverse to restore every scratch qubit to $\ket{0}$ and the input register to $\ket{x}$.
\end{itemize}

\end{document}

%% file: GH_notes/dfthc_integrals_optimization_target.tex
Let $[N]\coloneqq\{0,\ldots,N-1\}$ and consider the real spin-free
electronic-structure Hamiltonian
\begin{equation}\label{eq:chem_ham}
    H
    =
    \sum_{p,q\in[N]} h^{(1)}_{pq}E_{pq}
    +
    \frac{1}{2}
    \sum_{p,q,r,s\in[N]}
    h^{(2)}_{pqrs}E_{pq}E_{rs},
    \qquad
    E_{pq}\coloneqq
    \sum_{\sigma\in\{0,1\}}a^\dagger_{p\sigma}a_{q\sigma}.
\end{equation}
For our work we will search for representations of the Hamiltonian factorized into a form amenable to efficient circuit implementation. As such we can consider a related class of Hamiltonians that is spectrally equivalent in an $\eta$-particle manifold of Hilbert space
\begin{align}
    H
    =
    \sum_{p,q\in[N]} \left(h^{(1)}_{pq} - \frac{\eta}{2}S_{pq} + \beta_{1}\delta_{pq}\right)E_{pq}
    +
    \frac{1}{2}
    \sum_{p,q,r,s\in[N]}
    \left(h^{(2)}_{pqrs} + \frac{1}{2}\left(S \otimes \mathbb{I}_{N} + \mathbb{I}_{N} \otimes S\right)_{pqrs} \right)E_{pq}E_{rs} + \left(-\eta\beta_{1}\right)\mathbb{I}_{4^{N}},
\end{align}
for a symmetric matrix $S$ and scalar $\beta_{1}$.
In the competition, the symmetry shift $(S, \beta_{1})$ is not an optimization variable: the verifier scores every submission with the shift of the provided baseline factorization, and any submitted shift is ignored.
The inputs to the factorization problem are
\begin{equation}
    \left(
        h^{(1)},h^{(2)},R,B,C,\mathcal H_{\mathrm{sym}},
        \epsilon_{\mathrm{DFTHC}}
    \right),
    \qquad
    \epsilon_{\mathrm{DFTHC}}
    \coloneqq
    20\,\mathrm{mHa}
    =
    2.0\times 10^{-2}\,\mathrm{Ha},
\end{equation}
where $\mathcal H_{\mathrm{sym}}$ is the particle-number ($\eta$)
in which the ground-state energy is to be estimated.  The integers
$R$, $B$, and $C$ respectively specify the number of rank components,
the number of single-particle vectors per component, and the number of
coefficient copies per component of the following tensor network description of the two-electron integrals $h_{pqrs}^{(2)}$.

For every $\rho\in[R]$ and $b\in[B]$, introduce a real unit vector
\begin{equation}
    u^{(\rho)}_b\in\mathbb R^N,
    \qquad
    \bigl\|u^{(\rho)}_b\bigr\|_2=1,
\end{equation}
and, for every $(\rho,c)\in[R]\times[C]$, introduce real coefficients
$w_b^{(\rho c)}$ and an identity offset $W^{(\rho c)}$.  Define
\begin{equation}
    X^{(\rho c)}_{pq}
    \coloneqq
    \sum_{b\in[B]}
    w_b^{(\rho c)}
    u^{(\rho)}_{b,p}u^{(\rho)}_{b,q},
    \qquad
    \widetilde h^{(2)}_{pqrs}
    \coloneqq
    \sum_{\rho\in[R]}\sum_{c\in[C]}
    X^{(\rho c)}_{pq}X^{(\rho c)}_{rs}.
\end{equation}
The corresponding DFTHC Hamiltonian is
\begin{align}
    H_{\mathrm{DFTHC}}(\Theta)
    &\coloneqq
    \sum_{p,q}h^{(1)\prime}_{pq}E_{pq}
    +
    \frac{1}{2}
    \sum_{\rho,c}
    \left(
        W^{(\rho c)}\mathbb{I}_{4^{N}}
        +
        \sum_{p,q}X^{(\rho c)}_{pq}E_{pq}
    \right)^2
    -\left(\eta \beta_{1} + 
    \frac{1}{2}\sum_{\rho,c}\bigl|W^{(\rho c)}\bigr|^2 \right)\mathbb{I}_{4^{N}},
    \label{eq:dfthc-hamiltonian}
    \\
    \Theta
    &\coloneqq
    \left\{
        u_b^{(\rho)},w_b^{(\rho c)},W^{(\rho c)}, h_{\text{sym}}^{(1)}, \beta_{1}
    \right\}_{\rho,b,c}
\end{align}
where $h_{\text{sym}}^{(1)}$ is a real symmetric matrix and $\beta_{1}$ is a real-valued scalar.
The free one-body tensor is chosen as
\begin{equation}
    h^{(1)\prime}_{pq}
    \coloneqq
    h^{(1)}_{pq}
    -
    \frac{\eta}{2}S_{pq} + \beta_{1}\delta_{pq} -
    \sum_{\rho,c}
    W^{(\rho c)}X^{(\rho c)}_{pq}.
    \label{eq:exact-one-body-absorption}
\end{equation}
Consequently, expanding Eq.~\eqref{eq:dfthc-hamiltonian} gives
\begin{equation}
    H_{\mathrm{DFTHC}}(\Theta)
    =
    -\eta\beta_{1} \mathbb{I}_{4^{N}} + 
    \sum_{p,q}\left(h^{(1)}_{pq} - \frac{\eta}{2}S_{pq} + \beta_{1}\delta_{pq}\right)E_{pq}
    +
    \frac{1}{2}
    \sum_{p,q,r,s}
    \widetilde h^{(2)}_{pqrs}E_{pq}E_{rs}.
\end{equation}
Thus the one-body interaction is represented exactly for every
$\Theta$; only the two-body tensor is approximated.  Its error is
measured by
\begin{equation}
    \epsilon_{\mathrm{fro}}(\Theta)
    \coloneqq
    \left\|
        h^{(2)}-\widetilde h^{(2)}(\Theta)
    \right\|_{\mathrm F}
    =
    \left(
        \sum_{p,q,r,s}
        \left|
            h^{(2)}_{pqrs} 
            -
            \widetilde h^{(2)}_{pqrs}(\Theta) + \frac{1}{2}\left(S \otimes \mathbb{I}_{N} + \mathbb{I}_{N} \otimes S\right)_{pqrs}
        \right|^2
    \right)^{1/2}.
\end{equation}
We will also check if the residual of the $1$-electron component of the Hamiltonian is supplied correctly.  

To define the quantum-cost objective, let
\begin{equation}
    w_B^{(\rho c)}
    \coloneqq
    W^{(\rho c)}
    +
    \sum_{b\in[B]}w_b^{(\rho c)}.
\end{equation}
The LCU one-norm, equivalently the normalization of the DFTHC
Hamiltonian block encoding, is
\begin{equation}
    \Lambda(\Theta)
    \coloneqq
    \bigl\|h^{(1)\prime}\bigr\|_*
    +
    \frac{1}{4}
    \sum_{\rho,c}
    \left(
        \bigl|w_B^{(\rho c)}\bigr|
        +
        \sum_{b\in[B]}\bigl|w_b^{(\rho c)}\bigr|
    \right)^2,
    \label{eq:dfthc-normalization}
\end{equation}
where $\|\cdot\|_*$ denotes the Schatten one-norm.  The associated
sum-of-squares energy shift is
\begin{equation}
    E_{\mathrm{SOS}}(\Theta)
    \coloneqq 
    -\eta \beta_{1}
    -
    \left(
        \bigl\|h^{(1)\prime}\bigr\|_*
        -
        \operatorname{Tr}h^{(1)\prime}
    \right)
    -
    \frac{1}{2}
    \sum_{\rho,c}\bigl|W^{(\rho c)}\bigr|^2,
\end{equation}
so that
$H_{\mathrm{DFTHC}}(\Theta)-E_{\mathrm{SOS}}(\Theta)I\succeq0$.
Define
\begin{align}
    E_{\mathrm{gs}}(\Theta)
    &\coloneqq
    \lambda_{\min}
    \left(
        H_{\mathrm{DFTHC}}(\Theta)
        \big|_{\mathcal H_{\mathrm{sym}}}
    \right),
    \\
    E_{\mathrm{gap}}(\Theta)
    &\coloneqq
    E_{\mathrm{gs}}(\Theta)-E_{\mathrm{SOS}}(\Theta),
    \qquad
    0\leq E_{\mathrm{gap}}(\Theta)\leq 2\Lambda(\Theta).
\end{align}
Writing
\begin{equation}
    \delta_{\mathrm{gap}}(\Theta)
    \coloneqq
    \frac{E_{\mathrm{gap}}(\Theta)}{\Lambda(\Theta)},
    \qquad
    g_{\mathrm{amp}}(\Theta)
    \coloneqq
    \sqrt{
        \delta_{\mathrm{gap}}(\Theta)
        \bigl(2-\delta_{\mathrm{gap}}(\Theta)\bigr)
    },
\end{equation}
the spectrum-amplified block-encoding one-norm is the product
\begin{equation}
    \lambda_{\mathrm{eff}}(\Theta)
    \coloneqq
    \Lambda(\Theta)g_{\mathrm{amp}}(\Theta)
    =
    \sqrt{
        E_{\mathrm{gap}}(\Theta)
        \left(
            2\Lambda(\Theta)-E_{\mathrm{gap}}(\Theta)
        \right)
    }.
    \label{eq:effective-normalization}
\end{equation}

For fixed $(R,B,C)$, a good DFTHC factorization is therefore any
solution of the constrained optimization problem
\begin{equation}
    \boxed{
    \begin{aligned}
        \Theta^\star
        \in\;
        &\underset{\Theta}{\operatorname{argmin}}
        &&
        \lambda_{\mathrm{eff}}(\Theta)
        \\[1mm]
        &\text{subject to}
        &&
        \epsilon_{\mathrm{fro}}(\Theta)
        <
        \epsilon_{\mathrm{DFTHC}}
        =
        20\,\mathrm{mHa},
        \\
        &&
        &\bigl\|u_b^{(\rho)}\bigr\|_2=1
        \quad
        \text{for all }\rho\in[R],\ b\in[B].
    \end{aligned}}
    \label{eq:dfthc-optimization}
\end{equation}
The output consists of $\Theta^\star$, the exactly absorbed
one-body tensor $h^{(1)\prime}(\Theta^\star)$, and the diagnostics
$\epsilon_{\mathrm{fro}}$, $\Lambda$, $E_{\mathrm{SOS}}$,
$E_{\mathrm{gap}}$, and $\lambda_{\mathrm{eff}}$.  If different
choices of $(R,B,C)$ or different block-encoding circuit
implementations are compared, the implementation-aware outer
objective is
\begin{equation}
    \mathcal C_{\mathrm{BE}}(R,B,C)\,
    \lambda_{\mathrm{eff}}(\Theta^\star),
\end{equation}
where $\mathcal C_{\mathrm{BE}}$ is the cost of one query to the
selected DFTHC block-encoding circuit.

%% file: GH_notes/majorana_rotations_classical.tex
\setlength{\fboxsep}{6pt}
\begin{equation}\label{eq:tree_majorana_coefficients}
\fbox{%
\begin{minipage}{0.88\linewidth}
\small
\textbf{Black-box routine:}
\(\mathbf u\leftarrow\operatorname{TreeMajoranaCoefficients}(\mathcal T,\vec\theta)\).\par\smallskip
\textbf{Input:} A rooted, child-ordered tree \(\mathcal T\) on \(N\) vertices and an edge-angle vector
\(\vec\theta\in\mathbb R^{N-1}\), listed in the parent-before-child order encoded by \(\mathcal T\).\par\smallskip
\textbf{Output:} The coefficient vector \(\mathbf u(\vec\theta)\in\mathbb R^N\), independent of \(x\in\{0,1\}\), defining
\(\gamma_{\mathbf u(\vec\theta),x}:=\sum_vu_v(\vec\theta)\gamma_{v,x}\) and satisfying
\(U_{\mathcal T}(\vec\theta)^\dagger\gamma_{0,x}U_{\mathcal T}(\vec\theta)=\gamma_{\mathbf u(\vec\theta),x}\).\par\smallskip
\textbf{Classical algorithm runtime:} \(O(N)\) floating-point operations and \(O(N)\) output storage; see the algorithm below.
\end{minipage}%
}
\end{equation}

Let \(\mathcal T=(V,E)\) be a tree on \(V=\{0,\ldots,N-1\}\), rooted at vertex \(0\).  For each orbital \(v\), let \(\gamma_{v,x}\), \(x\in\{0,1\}\), denote its two Majorana operators, with
\[
    \{\gamma_{v,x},\gamma_{w,y}\}=2\delta_{vw}\delta_{xy}I.
\]
Orient every edge away from the root, fix an order on the children of each vertex, and enumerate the edges in any parent-before-child order
\[
    e_k=(p_k\to q_k),\qquad k=1,\ldots,N-1,
\]
meaning that the edge entering a nonroot vertex precedes every edge leaving that vertex.  Associate angle \(\theta_k\) with \(e_k\).  The sibling order is part of the specification because rotations on incident edges need not commute.

For an oriented edge \((v\to w)\), define the fermionic Givens rotation
\[
    G_{vw}(\theta):=
    \exp\!\left[\frac{\theta}{2}
    \bigl(\gamma_{v,0}\gamma_{w,0}+\gamma_{v,1}\gamma_{w,1}\bigr)\right],
\]
with convention
\[
    G_{vw}(\theta)^\dagger
    \begin{pmatrix}\gamma_{v,x}\\ \gamma_{w,x}\end{pmatrix}
    G_{vw}(\theta)
    =
    \begin{pmatrix}\cos\theta&\sin\theta\\-\sin\theta&\cos\theta\end{pmatrix}
    \begin{pmatrix}\gamma_{v,x}\\ \gamma_{w,x}\end{pmatrix},
    \qquad x\in\{0,1\}.
\]
Set
\[
    U_{\mathcal T}(\vec\theta)
    :=\prod_{k=1}^{N-1}G_{p_kq_k}(\theta_k),
    \qquad\text{with factors ordered from left to right in increasing }k.
\]
Then there is a real unit vector \(\mathbf u(\vec\theta)\in\mathbb R^N\) such that, simultaneously for \(x=0,1\),
\begin{equation}\label{eq:tree_unit_vec}
    U_{\mathcal T}(\vec\theta)^\dagger\gamma_{0,x}U_{\mathcal T}(\vec\theta)
    =\sum_{v=0}^{N-1}u_v(\vec\theta)\gamma_{v,x}
    =:\gamma_{\mathbf u(\vec\theta),x},
    \qquad \sum_{v=0}^{N-1}u_v^2=1.
\end{equation}
Thus a coherent implementation uses \(N-1\) angle registers, one for each edge, and applies the corresponding controlled Givens rotations in the specified order.

\paragraph{Classical coefficient algorithm.}
The coefficients in Eq.~\eqref{eq:tree_unit_vec} are computed by applying the same planar rotations to \(\mathbf e_0\):
\begin{equation}\label{eq:tree_planar_alg}
\begin{array}{ll}
\textbf{Input:} & (p_k,q_k,\theta_k)_{k=1}^{N-1}.\\
\textbf{Initialize:} & \mathbf u\leftarrow(1,0,\ldots,0).\\
\textbf{For }k=1,\ldots,N-1: &
\begin{aligned}[t]
    c&\leftarrow\cos\theta_k,\qquad s\leftarrow\sin\theta_k,\\
    a&\leftarrow u_{p_k},\qquad b\leftarrow u_{q_k},\\
    u_{p_k}&\leftarrow ca-sb,\qquad
    u_{q_k}\leftarrow sa+cb.
\end{aligned}\\
\textbf{Return:} & \mathbf u.
\end{array}
\end{equation}
The ordering condition implies that \(q_k\) has not been touched before its parent edge is processed, so \(u_{q_k}=0\) at step \(k\).  Hence the update simplifies to
\begin{equation}\label{eq:tree_simplified_update}
    u_{q_k}\leftarrow u_{p_k}\sin\theta_k,
    \qquad
    u_{p_k}\leftarrow u_{p_k}\cos\theta_k.
\end{equation}
Each of the \(N-1\) edges is processed once.  The algorithm therefore uses \(O(N)\) floating-point operations and \(O(N)\) output storage; after evaluating the trigonometric functions, Eq.~\eqref{eq:tree_simplified_update} uses exactly \(2(N-1)\) multiplications.  Each update is orthogonal, so it preserves \(\|\mathbf u\|_2=1\).

Equivalently, if the ordered children of \(v\) are \(w_1,\ldots,w_{d_v}\) and \(A_v\) is the coefficient arriving at \(v\), then
\begin{equation}\label{eq:tree_child_product}
    A_0=1,\qquad
    A_{w_j}=A_v\!\left(\prod_{\ell<j}\cos\theta_{vw_\ell}\right)\sin\theta_{vw_j},
    \qquad
    u_v=A_v\prod_{\ell=1}^{d_v}\cos\theta_{vw_\ell}.
\end{equation}
For the path \(0\to1\to\cdots\to N-1\), this reduces to the usual sequential line construction.

%% file: GH_notes/dfthc_block_encoding.tex
Throughout, $[M]\coloneqq\idx{M}$.  For a batch size $1\le J\le N-1$, define $K\coloneqq\lceil(N-1)/J\rceil$.  For optimization, the DFTHC block-encoding circuit is specified by a fixed tree $\mathcal T$, the black-box unitaries \OUTER{} and \INNER, and a family of Boolean angle oracles $\{\ANGLE_k\}_{k=0}^{K-1}$.  The first two are supplied as black boxes and enter the objective only through heuristic Toffoli costs.  Only the angle-oracle circuits are synthesized and optimized explicitly.  All generator labels, register maps, spin controls, Majorana signs, and reflection details are hidden from the user-facing interface and deferred to~\cref{app:dfthc_circuits}.

Let $\mathcal T=(V,E)$ be a rooted, child-ordered tree on $V=[N]$.  Its oriented edges are listed as $e_t=(p_t\to q_t)$, $t\in[N-1]$, in parent-before-child order.  For every opaque unit-vector address $a$, let $\vec\theta_a=(\theta_{a,0},\ldots,\theta_{a,N-2})$ be the edge angles represented by the family $\{\ANGLE_k\}_{k=0}^{K-1}$ and define
\[
  \mathbf u_a
  \coloneqq \TreeCoeff(\mathcal T,\vec\theta_a)\in\mathbb R^N,
  \qquad
  U_{\mathcal T,\sigma}(\vec\theta_a)^\dagger
  \gamma_{0\sigma x}
  U_{\mathcal T,\sigma}(\vec\theta_a)
  =\sum_{p\in[N]}u_{a,p}\gamma_{p\sigma x}.
\]
The same real unit vector is obtained for both $x\in\{0,1\}$ and either spin $\sigma$.  The classical map uses $O(N)$ floating-point operations and $O(N)$ working storage.  Its explicit recursion is given in~\cref{eq:tree-coefficient-recursion}.

Let $\ell\in[L]$, where $L=RC$, be an opaque label for a two-electron square.  The $L$ squares are internally grouped into $R$ sets of $C$ copies, with the same $B$ vectors shared inside each group; this bookkeeping is hidden from the user-facing problem.  For $b\in[B]$, write $w_{\ell b}\in\mathbb R$, and let $u_{\ell b}$ denote the addressed vector returned by $\TreeCoeff$.  Every one-body eigenvector used below is represented in the same way.  Let $W_\ell$ be the identity offset and define $w_{\ell B}=W_\ell+\sum_{b\in[B]}w_{\ell b}$.  The implemented tensors are specified classically by
\begin{equation}
\begin{aligned}
  X^{(\ell)}_{pq}
  &\coloneqq \sum_{b\in[B]}w_{\ell b}
       u_{\ell b,p}u_{\ell b,q},
  &
  \widetilde h^{(2)}_{pqrs}
  &\coloneqq \sum_{\ell\in[L]}X^{(\ell)}_{pq}X^{(\ell)}_{rs},
  \\
  h^{(1)\prime}_{pq}
  &\coloneqq h^{(1)}_{pq}-\frac{\eta}{2}S_{pq} + \beta_{1}\delta_{pq} - \sum_{\ell\in[L]}W_\ell X^{(\ell)}_{pq}.
\end{aligned}
\label{eq:classical-data}
\end{equation}
Thus the one-electron tensor is absorbed exactly (though still verified); only
$\widetilde h^{(2)}$ approximates the input tensor $h^{(2)}$.

\begin{tcolorbox}[
  title={\textbf{Black-box input and block-encoding output}},
  breakable,
  colback=black!2,colframe=black!65,
  boxrule=0.7pt,arc=1.5pt,
  left=7pt,right=7pt,top=5pt,bottom=5pt]
\textbf{Inputs.}  Integers $N,R,B,C$; target tensors $h^{(1)}$ and $h^{(2)}$; weights $\{w_{\ell b},W_\ell\}$; the rooted, child-ordered tree $\mathcal T$; an angle precision $b_{\mathrm{rot}}$; a batch size $1\le J\le N-1$; and the following unitaries.

\begin{equation}
  \OUTER\ket{0}
  =\sum_{x\in\mathcal X}\alpha_x\ket{x},
  \qquad
  \Tcost(\OUTER)\coloneqq C_{\mathrm{out}}.
\label{eq:outer}
\end{equation}
Here $x$ is an opaque outer address.  Its amplitudes encode the relative
normalizations of the one- and two-electron square-root generators.

\begin{equation}
  \INNER:\quad
  \ket{x}\ket{0}
  \longmapsto
  \ket{x}\sum_{y\in\mathcal Y_x}\beta_{x,y}\ket{y},
  \qquad
  \Tcost(\INNER)\coloneqq C_{\mathrm{in}}.
\label{eq:inner}
\end{equation}
Here $y$ is an opaque inner address.  It selects the identity or rotated
Majorana contribution and includes any required sign, phase, success, and
garbage registers.

For $k\in[K]$, let
\[
\begin{aligned}
  \bm\vartheta_a^{[k]}
  &\coloneqq
  \bigl(\vartheta_{a,kJ},\ldots,\vartheta_{a,kJ+J-1}\bigr),
  \\
  \Delta\bm\vartheta_a^{[k]}
  &\coloneqq
  \begin{cases}
    \bm\vartheta_a^{[0]}, & k=0,\\
    \bm\vartheta_a^{[k]}\oplus\bm\vartheta_a^{[k-1]}, & k>0,
  \end{cases}
\end{aligned}
\]
where $\vartheta_{a,t}\in\{0,1\}^{b_{\mathrm{rot}}}$ encodes $\theta_{a,t}/\pi$ and is zero for $t\ge N-1$.  The batch label is hard-wired into the oracle: there is no coherent $k$ register, and the only input/control is the vector address $a$,
\begin{equation}
  \ANGLE_k:\quad
  \ket a\ket{\bm z}
  \longmapsto
  \ket a\ket{\bm z\oplus\Delta\bm\vartheta_a^{[k]}},
  \qquad
  \Tcost(\ANGLE_k)\coloneqq C_{\mathrm{ang},k}(J),
  \quad k\in[K].
\label{eq:angle}
\end{equation}
XOR is componentwise over the $J$ angle words.  Thus, for $k>0$, a register holding $\bm\vartheta_a^{[k-1]}$ is changed directly into $\bm\vartheta_a^{[k]}$, eliminating a separate inter-batch uncomputation.  One forward traversal costs
$C_{\mathrm{ang}}^{\mathrm{pass}}(J)\coloneqq\sum_{k\in[K]}C_{\mathrm{ang},k}(J)$.

\textbf{Output.}  A unitary $U_{\mathrm{BE}}$ and signal projector $\Pi$
satisfying
\begin{equation}
  \Pi U_{\mathrm{BE}}\Pi
  =
  \left(
    \frac{H_{\mathrm{DFTHC}}-E_{\mathrm{SOS}}I}{\Lambda}-I
  \right)\Pi.
\label{eq:output}
\end{equation}
The tree and these three oracle interfaces completely specify the implemented one- and two-electron block encoding.  Each angle oracle is self-inverse, and each black-box inverse has the same assigned Toffoli cost.
\end{tcolorbox}

\begin{tcolorbox}[
  colback=blue!2,colframe=blue!45!black,
  boxrule=0.65pt,arc=1.5pt,
  left=7pt,right=7pt,top=4pt,bottom=4pt]
\textbf{Optimization scope.}
$\OUTER$ and $\INNER$ are never opened or re-synthesized: their contributions are represented only by heuristic values $C_{\mathrm{out}}$ and $C_{\mathrm{in}}$.  The tree $\mathcal T$ is fixed classical structure, and $\TreeCoeff$ is a deterministic classical evaluation map.  The reversible implementations and scheduling of $\{\ANGLE_k\}_{k=0}^{K-1}$ are the only circuit-level optimization variables; their explicit costs are $C_{\mathrm{ang},k}(J)$, or $C_{\mathrm{ang}}^{\mathrm{pass}}(J)$ for one complete traversal.
\end{tcolorbox}

\Needspace{5\baselineskip}
\paragraph{Streamed, batched, or parallel angles.}
Starting from $\bm z=0$, $\ANGLE_0$ loads batch $0$; after batch $k-1$ is consumed, $\ANGLE_k$ overwrites it in place with batch $k$.  The reverse traversal walks through the batches backward and ends with $\ANGLE_0$, which clears $\bm z$.  Hence $J=1$ streams one angle, intermediate $J$ reuses $J$ angle-word registers, and $J=N-1$ loads all angles at once.  The choice of $J$ changes ancillas, latency, and cost, but not the represented Hamiltonian.

\subsubsection{Verification and scoring}

For FeMo-co ($N=76$, $\eta=113$), the verifier fixes
$S=S_0$ and $\beta_1=\beta_{1,0}$ to the supplied baseline.
Participants may optimize the factorization ranks and factors,
angle precision, and angle-loading circuits.

The verifier simulates every valid vector address, with 64
measurement samples per address, checking consistent angle words,
unchanged addresses, clean workspace, and corrected phases during
forward and reverse traversal. Each angle circuit is limited to
128 qubits. The recovered words determine the vectors used in
all subsequent calculations.

Using these vectors, the verifier constructs $X^{(\rho c)}$ and
$h^{(1)\prime}$ as defined above. It matches the decoded one-body
vectors to the eigenvectors of $h^{(1)\prime}$ and reconstructs
$\widetilde h^{(1)\prime}$ using the corresponding eigenvalues.
Acceptance requires
\begin{align}
\epsilon_1
&=\|\widetilde h^{(1)\prime}-h^{(1)\prime}\|_{\mathrm F}
\le 0.020\,\mathrm{Ha},\\
\epsilon_2
&=\left\|\sum_{\rho,c}X^{(\rho c)}\otimes X^{(\rho c)}
-h^{(2)}-\mathcal T(S_0)\right\|_{\mathrm F}
\le 0.020\,\mathrm{Ha},
\end{align}
where $\mathcal T(S_0)=(S_0\otimes I+I\otimes S_0)/2$.

The verifier recomputes $\Lambda$ and $E_{\mathrm{SOS}}$ and uses
\begin{align}
E_{\mathrm{gap}}&=E_{\mathrm{upper}}-E_{\mathrm{SOS}},\\
\lambda_{\mathrm{eff}}
&=\sqrt{E_{\mathrm{gap}}(2\Lambda-E_{\mathrm{gap}})},
\end{align}
with the fixed, core-inclusive reference
$E_{\mathrm{upper}}=-22140.27335\,\mathrm{Ha}$.
All metrics must be finite, with
$0<E_{\mathrm{gap}}<2\Lambda$,
$E_{\mathrm{gap}} > 3.914898\,\mathrm{Ha}$, and
$\Lambda > 49.0\,\mathrm{Ha}$.

Accepted submissions receive the score
\begin{equation}
\mathcal S=\lambda_{\mathrm{eff}}T_{\mathrm{tot}}Q_{\mathrm{tot}},
\end{equation}
using simulated angle-circuit costs and the prescribed
state-preparation and Givens-rotation cost model.
$Q_{\mathrm{tot}}$ includes the full modeled workspace and is
not capped at 128. Failed checks give infinite score; values
are not clamped to the floors.

The original baseline passes with unchanged score.
These checks constrain coefficient errors and resource costs;
they do not certify a 20 mHa many-electron energy error.
\subsection*{Classical quantities exposed to the user}

Before evaluating either quantity below, the user decodes the angle words and applies $\TreeCoeff(\mathcal T,\vec\theta_a)$ to each addressed vector.  For the $N+RB$ distinct vectors used by the standard DFTHC organization, this preprocessing costs $O((N+RB)N)$ floating-point operations and can be streamed with $O(N)$ working storage.  It is lower order than the tensor contractions below.

\paragraph{Two-electron Frobenius error.}
The implemented error is
\begin{equation}
  \epsilon_{\mathrm{fro}}
  \coloneqq
  \left\|h^{(2)}-\widetilde h^{(2)}\right\|_{\mathrm F}
  =
  \left[
    \sum_{p,q,r,s}
    \left(
      h^{(2)}_{pqrs}
      -\sum_{\ell}X^{(\ell)}_{pq}X^{(\ell)}_{rs}
    \right)^2
  \right]^{1/2}.
\label{eq:fro-direct}
\end{equation}
It need not be evaluated by explicitly storing $\widetilde h^{(2)}$.  Let
$\mathsf H_2\in\mathbb R^{N^2\times N^2}$ be the pair-index flattening
$(\mathsf H_2)_{(pq),(rs)}=h^{(2)}_{pqrs}$ and let
$x_\ell=\operatorname{vec}(X^{(\ell)})$.  Then
\begin{equation}
  \epsilon_{\mathrm{fro}}^2
  =\left\|h^{(2)}\right\|_{\mathrm F}^2
   -2\sum_{\ell}x_\ell^{\mathsf T}\mathsf H_2x_\ell
   +\sum_{\ell,m}\left(x_\ell^{\mathsf T}x_m\right)^2.
\label{eq:fro-contraction}
\end{equation}
With ordinary dense floating-point contractions this costs
\begin{equation}
  O\!\left(
    N^4+LBN^2+LN^4+L^2N^2
  \right)
\label{eq:fro-cost}
\end{equation}
operations, with dominant contraction cost $O(LN^4)=O(RCN^4)$.  Under the
circuit-design scaling $B,C=\Theta(N)$ and $R=\widetilde O(1)$, the exponent
is \textbf{$5$}: $\epsilon_{\mathrm{fro}}$ is computable in
$\widetilde O(N^5)$ dense floating-point operations (or $O(N^5)$ when $R$ is
constant).
\paragraph{1-electron Frobenius norm error}
 We also check the loss with respect to the correct $1$-electron coefficients. This is computed as a function of the tensors supplied by the user $S$, $W^{\rho, c}$, $X_{pq}^{\rho c}$. This is to ensure that the QROM output provided for the 1-electron component of the Hamiltonian is sufficiently low error to simulate the true $H_{\text{DFTHC}}$ Hamiltonian.  Recall the first $N$ vectors output from the collection of AngleK QROMs correspond to the eigenvectors of
 \begin{align}
h^{(1)\prime}_{pq} = h^{(1)}_{pq}-\frac{\eta}{2}S_{pq} + \beta_{1}\delta_{pq} - \sum_{\ell\in[L]}W_\ell X^{(\ell)}_{pq}
 \end{align}
 where $X^{(\ell)}_{pq}$ is computed from the output of the last $RB$ vectors of the AngleK QROMs.  Ultimately we want to check if 
 \begin{align}
\epsilon_{1,fro} = ||h^{(1)}_{pq} - \tilde{h}^{(1)}_{pq}||
 \end{align}
 is small and below the given threshold of 20 mHa. To do this we first construct $h^{(1)\prime}$ and diagonalize to get eigenvalues.  Next we take the eigenvectors exported from QROM and align them with the eigenvectors used for diagonalization in the previous step. We then reconstruct 
 \begin{align}
\tilde{h}^{(1)\prime} = \sum_{j}\lambda_{j}\tilde{v}_{j}\tilde{v}_{j}^{\dagger}
 \end{align}
 where $\lambda_{j}$ are the ideal eigenvalues and $\tilde{v}_{j}$ are the vectors read from QROM where elements are represented in $\beta$ bits of precision. Finally we recover $\tilde{h}^{(1)}$ by
 \begin{align}
\tilde{h}^{(1)}_{pq} = \tilde{h}^{(1)\prime}_{pq} + \frac{\eta}{2}S_{pq} - \beta_{1} \delta_{pq} + \sum_{\rho,c}W^{\rho,c}\tilde{X}_{pq}^{\rho c}
 \end{align}
 which enables the $\epsilon_{1,fro}$ computation. 
\paragraph{Block-encoding normalization.}
Diagonalize the exactly corrected one-body matrix as
\begin{equation}
  h^{(1)\prime}
  =\sum_j w_{+,j}u_{+,j}u_{+,j}^{\mathsf T}
   -\sum_j w_{-,j}u_{-,j}u_{-,j}^{\mathsf T},
  \qquad w_{\pm,j}\geq 0.
\end{equation}
The normalization is the sum of its one- and two-electron contributions:
\begin{equation}
\boxed{
  \Lambda
  =\underbrace{\sum_j\bigl(w_{+,j}+w_{-,j}\bigr)}_{\displaystyle
      \Lambda_1=\|h^{(1)\prime}\|_*}
   +\underbrace{\frac14\sum_{\ell\in[L]}
      \left(
        |w_{\ell B}|+\sum_{b\in[B]}|w_{\ell b}|
      \right)^2}_{\displaystyle\Lambda_2}.
}
\label{eq:lambda-main}
\end{equation}
Equations~\eqref{eq:fro-direct} and \eqref{eq:lambda-main} are the only
Hamiltonian-quality quantities required by the user-facing optimizer.

%% file: ref.bib
@article{BabbushSpectra,
    title = {{Encoding Electronic Spectra in Quantum Circuits with Linear T Complexity}},
    year = {2018},
    journal = {Physical Review X},
    author = {Babbush, Ryan and Gidney, Craig and Berry, Dominic and Wiebe, Nathan and McClean, Jarrod and Paler, Alexandru and Fowler, Austin and Neven, Hartmut},
    number = {4},
    pages = {041015},
    volume = {8},
    url = {https://journals.aps.org/prx/abstract/10.1103/PhysRevX.8.041015}
}

@article{Lee2020,
    title = {{Even More Efficient Quantum Computations of Chemistry Through Tensor Hypercontraction}},
    year = {2021},
    journal = {PRX Quantum},
    author = {Lee, Joonho and Berry, Dominic W. and Gidney, Craig and Huggins, William J. and McClean, Jarrod R. and Wiebe, Nathan and Babbush, Ryan},
    number = {3},
    month = {7},
    pages = {030305},
    volume = {2},
    publisher = {American Physical Society},
    url = {https://link.aps.org/doi/10.1103/PRXQuantum.2.030305},
    doi = {10.1103/PRXQuantum.2.030305},
    issn = {2691-3399}
}

@article{Su2021,
    title = {{Fault-tolerant quantum simulations of chemistry in first quantization}},
    year = {2021},
    journal = {PRX Quantum},
    author = {Su, Yuan and Berry, Dominic and Wiebe, Nathan and Rubin, Nicholas and Babbush, Ryan},
    number = {4},
    pages = {040332},
    volume = {2},
    url = {https://journals.aps.org/prxquantum/abstract/10.1103/PRXQuantum.2.040332}
}

@article{GidneyAdder,
    title = {{Halving the cost of quantum addition}},
    year = {2018},
    journal = {Quantum},
    author = {Gidney, Craig},
    pages = {74},
    volume = {2},
    url = {https://quantum-journal.org/papers/q-2018-06-18-74/},
    doi = {10.22331/q-2018-06-18-74}
}

@article{Gidney2019b,
    title = {{How to factor 2048 bit RSA integers in 8 hours using 20 million noisy qubits}},
    year = {2021},
    journal = {Quantum},
    author = {Gidney, Craig and Eker{\aa}, Martin},
    pages = {433},
    volume = {5},
    doi = {10.22331/q-2021-04-15-433},
    url = {https://quantum-journal.org/papers/q-2021-04-15-433/}
}

@article{Haner2018,
    title = {{Optimizing Quantum Circuits for Arithmetic}},
    year = {2018},
    journal = {arXiv:1805.12445},
    author = {H{\"{a}}ner, Thomas and Roetteler, Martin and Svore, Krysta M.},
    month = {5},
    url = {https://arxiv.org/abs/1805.12445}
}


%% file: references.bib
@article{rubin2024quantum,
  title={Quantum computation of stopping power for inertial fusion target design},
  author={Rubin, Nicholas C and Berry, Dominic W and Kononov, Alina and Malone, Fionn D and Khattar, Tanuj and White, Alec and Lee, Joonho and Neven, Hartmut and Babbush, Ryan and Baczewski, Andrew D},
  journal={Proceedings of the National Academy of Sciences},
  volume={121},
  number={23},
  pages={e2317772121},
  year={2024},
  publisher={National Academy of Sciences},
  url={https://www.pnas.org/doi/10.1073/pnas.2317772121}
}

@article{cody2012faster,
  title={Faster quantum chemistry simulation on fault-tolerant quantum computers},
  author={Jones, N. Cody and Whitfield, James D and McMahon, Peter L and Yung, Man-Hong and Van Meter, Rodney and Aspuru-Guzik, Al{\'a}n and Yamamoto, Yoshihisa},
  journal={New Journal of Physics},
  volume={14},
  number={11},
  pages={115023},
  year={2012},
  publisher={IOP Publishing},
  url={https://iopscience.iop.org/article/10.1088/1367-2630/14/11/115023}
}

@misc{harrigan2024qualtran,
    title={Expressing and Analyzing Quantum Algorithms with {Qualtran}},
    author={Matthew P. Harrigan and Tanuj Khattar
        and Charles Yuan and Anurudh Peduri and Noureldin Yosri
        and Fionn D. Malone and Ryan Babbush and Nicholas C. Rubin},
    year={2024},
    eprint={2409.04643},
    archivePrefix={arXiv},
    primaryClass={quant-ph},
    doi={10.48550/arXiv.2409.04643},
    url={https://arxiv.org/abs/2409.04643},
}

@article{pb2g-j9cw,
  title = {Fast Quantum Simulation of Electronic Structure by Spectrum Amplification},
  author = {Low, Guang Hao and King, Robbie and Berry, Dominic W. and Han, Qiushi and DePrince, III, A. Eugene and White, Alec F. and Babbush, Ryan and Somma, Rolando D. and Rubin, Nicholas C.},
  journal = {Phys. Rev. X},
  volume = {15},
  issue = {4},
  pages = {041016},
  numpages = {41},
  year = {2025},
  month = {Oct},
  publisher = {American Physical Society},
  doi = {10.1103/pb2g-j9cw},
  url = {https://link.aps.org/doi/10.1103/pb2g-j9cw}
}

@article{jordan2025dqi,
  title = {Optimization by decoded quantum interferometry},
  author = {Jordan, Stephen P. and Shutty, Noah and Wootters, Mary and
            Zalcman, Adam and Schmidhuber, Alexander and King, Robbie and
            Isakov, Sergei V. and Khattar, Tanuj and Babbush, Ryan},
  journal = {Nature},
  volume = {646},
  pages = {831--836},
  year = {2025},
  doi = {10.1038/s41586-025-09527-5},
  eprint = {2408.08292},
  archivePrefix = {arXiv},
  primaryClass = {quant-ph},
  url = {https://arxiv.org/abs/2408.08292v5}
}

@misc{khattar2025optimizeddqi,
      title={Verifiable Quantum Advantage via Optimized {DQI} Circuits}, 
      author={Tanuj Khattar and Noah Shutty and Craig Gidney and Adam Zalcman and Noureldin Yosri and Dmitri Maslov and Ryan Babbush and Stephen P. Jordan},
      year={2025},
      eprint={2510.10967},
      archivePrefix={arXiv},
      primaryClass={quant-ph},
      url={https://arxiv.org/abs/2510.10967}, 
}

@inproceedings{guruswami2000concatenated,
author = {Guruswami, Venkatesan and Sudan, Madhu},
title = {List decoding algorithms for certain concatenated codes},
year = {2000},
isbn = {1581131844},
publisher = {Association for Computing Machinery},
address = {New York, NY, USA},
url = {https://doi.org/10.1145/335305.335327},
doi = {10.1145/335305.335327},
booktitle = {Proceedings of the Thirty-Second Annual ACM Symposium on Theory of Computing},
pages = {181--190},
numpages = {10},
location = {Portland, Oregon, USA},
series = {STOC '00}
}

@inproceedings{naor1999oblivious,
  title = {Oblivious Transfer and Polynomial Evaluation},
  author = {Naor, Moni and Pinkas, Benny},
  booktitle = {Proceedings of the Thirty-First Annual ACM Symposium on
               Theory of Computing},
  pages = {245--254},
  year = {1999},
  publisher = {ACM},
  doi = {10.1145/301250.301312},
  url = {https://doi.org/10.1145/301250.301312}
}

@inproceedings{bleichenbacher2000noisy,
  title = {Noisy Polynomial Interpolation and Noisy {Chinese} Remaindering},
  author = {Bleichenbacher, Daniel and Nguyen, Phong Q.},
  booktitle = {Advances in Cryptology---EUROCRYPT 2000},
  series = {Lecture Notes in Computer Science},
  volume = {1807},
  pages = {53--69},
  year = {2000},
  publisher = {Springer},
  doi = {10.1007/3-540-45539-6_4},
  url = {https://www.iacr.org/archive/eurocrypt2000/1807/18070053-new.pdf}
}

@misc{proos2004shorsdiscretelogarithmquantum,
      title={Shor's discrete logarithm quantum algorithm for elliptic curves}, 
      author={John Proos and Christof Zalka},
      year={2004},
      eprint={quant-ph/0301141},
      archivePrefix={arXiv},
      primaryClass={quant-ph},
      url={https://arxiv.org/abs/quant-ph/0301141}, 
}

@inbook{fiat1986prove,
  title = {How To Prove Yourself: Practical Solutions to Identification and Signature Problems},
  ISBN = {9783540180470},
  url = {http://dx.doi.org/10.1007/3-540-47721-7_12},
  DOI = {10.1007/3-540-47721-7_12},
  booktitle = {Advances in Cryptology --- CRYPTO' 86},
  publisher = {Springer Berlin Heidelberg},
  author = {Fiat, Amos and Shamir, Adi},
  year = {1987},
  pages = {186--194}
}

@article{gidney2021stim,
  title = {Stim: a fast stabilizer circuit simulator},
  volume = {5},
  ISSN = {2521-327X},
  url = {http://dx.doi.org/10.22331/q-2021-07-06-497},
  DOI = {10.22331/q-2021-07-06-497},
  journal = {Quantum},
  publisher = {Verein zur Forderung des Open Access Publizierens in den Quantenwissenschaften},
  author = {Gidney, Craig},
  year = {2021},
  month = jul,
  pages = {497}
}

@misc{gidney2025factor2048,
  doi = {10.48550/ARXIV.2505.15917},
  url = {https://arxiv.org/abs/2505.15917},
  author = {Gidney, Craig},
  title = {How to factor 2048 bit {RSA} integers with less than a million noisy qubits},
  publisher = {arXiv},
  year = {2025},
  copyright = {Creative Commons Attribution 4.0 International}
}

@misc{babbush2026securingellipticcurvecryptocurrencies,
  doi = {10.48550/ARXIV.2603.28846},
  url = {https://arxiv.org/abs/2603.28846},
  author = {Babbush, Ryan and Zalcman, Adam and Gidney, Craig and Broughton, Michael and Khattar, Tanuj and Neven, Hartmut and Bergamaschi, Thiago and Drake, Justin and Boneh, Dan},
  title = {Securing Elliptic Curve Cryptocurrencies against Quantum Vulnerabilities: Resource Estimates and Mitigations},
  publisher = {arXiv},
  year = {2026},
  copyright = {Creative Commons Attribution 4.0 International}
}

@misc{long2026recordreplay,
  doi = {10.48550/ARXIV.2609.28882},
  url = {https://arxiv.org/abs/2609.28882},
  author = {Long, Jieyi and Pender, Theodore and Huang, Zhao and Santos, Manuel B. and Singh, Samrendra Kumar and Naskr{\k{e}}cki, Bartosz and {BitWonka} and Dallaire-Demers, Pierre-Luc and Giannicola, Francesco and Paschoarelli, Ruben M. L. and Freuler, Oli and Lee, Jackie Chia-Hsun and Gnuchev, Vasily and Kannappan, Gopi and Boyer, John and Butler, Xavier and Balasubramani, Akash and Newman, Jordan and Dereje, Bereket and Hertlein, Alexander and Kodra, Robert and Levy, Lucas and Patel, Shaan and Rose, JT and Zweil, Matt and Wisdom, Okechukwu and El-Eter, Tarek and Lee, Edison and Dong, Michael and Li, Alan and Joseph, Anto and Nguyen, Duy and Naik, Gajesh and Anant, Gautham and Deb, Soubhik and Drake, Justin},
  title = {Efficient Record-and-Replay Arithmetic for Quantum Elliptic-Curve Point Addition},
  publisher = {arXiv},
  year = {2026},
  copyright = {Creative Commons Attribution 4.0 International}
}

@article{janzing2005nonidentity,
  title = {``Non-identity-check'' is {QMA}-complete},
  volume = {03},
  ISSN = {1793-6918},
  url = {http://dx.doi.org/10.1142/S0219749905001067},
  DOI = {10.1142/s0219749905001067},
  number = {03},
  journal = {International Journal of Quantum Information},
  publisher = {World Scientific Pub Co Pte Ltd},
  author = {Janzing, Dominik and Wocjan, Pawel and Beth, Thomas},
  year = {2005},
  month = sep,
  pages = {463--473}
}

@misc{zhou2025neutralatom,
  doi = {10.48550/ARXIV.2505.15907},
  url = {https://arxiv.org/abs/2505.15907},
  author = {Zhou, Hengyun and Duckering, Casey and Zhao, Chen and Bluvstein, Dolev and Cain, Madelyn and Kubica, Aleksander and Wang, Sheng-Tao and Lukin, Mikhail D.},
  title = {Resource Analysis of Low-Overhead Transversal Architectures for Reconfigurable Atom Arrays},
  publisher = {arXiv},
  year = {2025},
  copyright = {arXiv.org perpetual, non-exclusive license}
}

@misc{cain2026shor,
  doi = {10.48550/ARXIV.2603.28627},
  url = {https://arxiv.org/abs/2603.28627},
  author = {Cain, Madelyn and Xu, Qian and King, Robbie and Picard, Lewis R. B. and Levine, Harry and Endres, Manuel and Preskill, John and Huang, Hsin-Yuan and Bluvstein, Dolev},
  title = {Shor's algorithm is possible with as few as 10,000 reconfigurable atomic qubits},
  publisher = {arXiv},
  year = {2026},
  copyright = {arXiv.org perpetual, non-exclusive license}
}

@inbook{deMoura2015,
  title = {The {Lean} Theorem Prover (System Description)},
  ISBN = {9783319214016},
  ISSN = {1611-3349},
  url = {http://dx.doi.org/10.1007/978-3-319-21401-6_26},
  DOI = {10.1007/978-3-319-21401-6_26},
  booktitle = {Automated Deduction - CADE-25},
  publisher = {Springer International Publishing},
  author = {de Moura,  Leonardo and Kong,  Soonho and Avigad,  Jeremy and van Doorn,  Floris and von Raumer,  Jakob},
  year = {2015},
  pages = {378--388}
}

@article{kliuchnikov2023shorter,
  title = {Shorter quantum circuits via single-qubit gate approximation},
  author = {Kliuchnikov, Vadym and Lauter, Kristin and Minko, Romy and Paetznick, Adam and Petit, Christophe},
  journal = {Quantum},
  volume = {7},
  pages = {1208},
  year = {2023},
  month = dec,
  publisher = {Verein zur F{\"{o}}rderung des Open Access Publizierens in den Quantenwissenschaften},
  doi = {10.22331/q-2023-12-18-1208},
  eprint = {2203.10064},
  archivePrefix = {arXiv},
  primaryClass = {quant-ph},
  url = {https://arxiv.org/abs/2203.10064}
}

@article{kissinger2020reducing,
  title = {Reducing the number of non-{Clifford} gates in quantum circuits},
  author = {Kissinger, Aleks and van de Wetering, John},
  journal = {Phys. Rev. A},
  volume = {102},
  issue = {2},
  pages = {022406},
  year = {2020},
  publisher = {American Physical Society},
  doi = {10.1103/PhysRevA.102.022406},
  url = {https://link.aps.org/doi/10.1103/PhysRevA.102.022406}
}

@inproceedings{xu2022quartz,
  title = {Quartz: superoptimization of quantum circuits},
  author = {Xu, Mingkuan and Li, Zikun and Padon, Oded and Lin, Sina and Pointing, Jessica and Hirth, Auguste and Ma, Henry and Palsberg, Jens and Aiken, Alex and Acar, Umut A. and Jia, Zhihao},
  booktitle = {Proceedings of the 43rd ACM SIGPLAN International Conference on Programming Language Design and Implementation},
  pages = {625--640},
  year = {2022},
  publisher = {ACM},
  doi = {10.1145/3519939.3523433},
  url = {https://doi.org/10.1145/3519939.3523433}
}

@article{ruiz2025alphatensor,
  title = {Quantum circuit optimization with {AlphaTensor}},
  author = {Ruiz, Francisco J. R. and Laakkonen, Tuomas and Bausch, Johannes and Balog, Matej and Barekatain, Mohammadamin and Heras, Francisco J. H. and Novikov, Alexander and Fitzpatrick, Nathan and Romera-Paredes, Bernardino and van de Wetering, John and Fawzi, Alhussein and Meichanetzidis, Konstantinos and Kohli, Pushmeet},
  journal = {Nature Machine Intelligence},
  volume = {7},
  pages = {374--385},
  year = {2025},
  doi = {10.1038/s42256-025-01001-1},
  url = {https://doi.org/10.1038/s42256-025-01001-1}
}

@misc{novikov2025alphaevolve,
  title = {{AlphaEvolve}: A coding agent for scientific and algorithmic discovery},
  author = {Novikov, Alexander and V{\~u}, Ng{\^a}n and Eisenberger, Marvin and Dupont, Emilien and Huang, Po-Sen and Wagner, Adam Zsolt and Shirobokov, Sergey and Kozlovskii, Borislav and Ruiz, Francisco J. R. and Mehrabian, Abbas and Kumar, M. Pawan and See, Abigail and Chaudhuri, Swarat and Holland, George and Davies, Alex and Nowozin, Sebastian and Kohli, Pushmeet and Balog, Matej},
  year = {2025},
  eprint = {2506.13131},
  archivePrefix = {arXiv},
  primaryClass = {cs.AI},
  url = {https://arxiv.org/abs/2506.13131}
}

@misc{gu2026discovering,
  title = {Discovering state-of-the-art quantum circuits with large language models},
  author = {Gu, Andi and Kemeny, Alek H. and Yelin, Susanne F. and Lukin, Mikhail D.},
  year = {2026},
  eprint = {2610.22367},
  archivePrefix = {arXiv},
  primaryClass = {quant-ph},
  url = {https://arxiv.org/abs/2610.22367}
}

@misc{babbush2026kitp,
  title = {Accelerating the field of quantum algorithms with {AI}},
  author = {Babbush, Ryan},
  howpublished = {Recorded talk at the {KITP} Conference: {Artificial Intelligence at the Quantum Frontier}},
  year = {2026},
  url = {https://www.youtube.com/watch?v=4rWyepeSUvM}
}

@misc{quake3_invsqrt,
  author = {{id Software}},
  title  = {{Quake III Arena} Source Code: \texttt{q\_math.c}},
  year   = {2005},
  url    = {https://github.com/id-Software/Quake-III-Arena/blob/dbe4ddb10315479fc00086f08e25d968b4b43c49/code/game/q_math.c},
  note   = {Commit \texttt{dbe4ddb}, lines 552--572}
}
